\documentclass[sigconf, 10pt]{acmart}

\usepackage{ifthen}
\usepackage{listings}
\usepackage{prp-macros}
\usepackage{booktabs}
\usepackage{subcaption}
\usepackage{url}
\usepackage{multicol}
\usepackage{multirow}
\usepackage{xcolor}
\usepackage{hhline}
\usepackage{listings}
\usepackage{enumitem}
\usepackage{xspace}
\usepackage{amssymb}
\usepackage{pifont}
\usepackage{framed}
\usepackage[vlined, linesnumbered]{algorithm2e}
\usepackage[skip=1ex]{caption}

\usepackage[disabled]{aplcomments} 

\lstnewenvironment{codelisting}{
	\lstset{
		language=[Sharp]C,
		basicstyle=\normalsize\ttfamily\linespread{0.8},
		tabsize=2,
		keywordstyle=\color{blue}\ttfamily,
		stringstyle=\color{red}\ttfamily,
		commentstyle=\color{magenta}\ttfamily,
		morecomment=[l][\color{magenta}]{\#},
		breaklines=true
	}
}{}

\SetAlFnt{\footnotesize\sffamily}

\SetCommentSty{mycommfont}
\newcommand{\eat}[1]{}

\newcommand{\DONE}[1]{}

\newcommand{\figref}[1]{Figure~\ref{fig:#1}}
\newcommand{\secref}[1]{Section~\ref{sec:#1}}

\newcommand{\ssecref}[1]{Sec.~\ref{sec:#1}}

\newcommand{\algref}[1]{Algorithm~\ref{alg:#1}}

\newcommand{\seclabel}[1]{\label{sec:#1}}
\newcommand{\figlabel}[1]{\label{fig:#1}}
\newcommand{\sseclabel}[1]{\label{sec:#1}}

\newcommand{\alglabel}[1]{\label{alg:#1}}

\newcommand{\ignore}[1]{}

\newcounter{programlinenumber}

\newboolean{TR}
\setboolean{TR}{false}
\ifthenelse{\boolean{TR}}{

\newcommand{\TrOnly}[1]{#1}
\newcommand{\SubOnly}[1]{}
\newcommand{\TrOnlyInFootnote}[1]{#1}
\newcommand{\TrOnlyInTable}[1]{#1}}
{

\newcommand{\TrOnly}[1]{}
\newcommand{\SubOnly}[1]{#1}
\newcommand{\TrOnlyInFootnote}[1]{}
\newcommand{\TrOnlyInTable}[1]{}}

\newcommand{\analysis}{\textsc{analysis}\xspace}
\newcommand{\conflictfree}{\textsc{conflict-free}\xspace}
\newcommand{\residual}{\textsc{residual}\xspace}
\newtheorem*{problemstatement*}{Problem Statement}
\newcommand{\preprocessing}{\textsc{pre-processing}\xspace}
\newcommand{\spot}{\textsc{spot}\xspace}
\newcommand{\allocate}{\textsc{allocate}\xspace}
\newcommand{\merge}{\textsc{merge}\xspace}

\newcommand{\set}[1]{\{#1\}}                    

\newcommand{\tpcc}{TPC-C\xspace}

\renewcommand\footnotetextcopyrightpermission[1]{} 

\setcopyright{none}
\acmDOI{10.475/123_4}

\acmISBN{123-4567-24-567/08/06}

\acmConference[SIGMOD'19]{ACM SIGMOD/PODS International Conference on Management of Data}{June 30- July 5, 2019}{Amsterdam, The Netherlands}
\acmYear{2019}
\copyrightyear{2019}

\graphicspath{images}
\begin{document}
\title[The Strife Transaction Scheduler]{Improving High Contention OLTP Performance via Transaction Scheduling}
\author{Guna Prasaad, Alvin Cheung, Dan Suciu\\
University of Washington\\
\texttt{\{guna, akcheung, suciu\}@cs.washington.edu}}

\renewcommand{\shortauthors}{G. Prasaad et al.}
\newcommand{\system}{\textsc{Strife}\xspace}
\newcommenter{guna}{1.0,0.4,1.0} 
\newcommenter{dan}{1.0,0.4,0.4} 
\newcommenter{alvin}{0.4,1.0,1.0} 

%
%



\begin{abstract}
Research in transaction processing has made significant progress in improving the performance of multi-core in-memory transactional systems.
However, the focus has mainly been on low-contention workloads. Modern transactional systems perform poorly on workloads with transactions accessing a few highly contended data items. 
We observe that most transactional workloads, including those with high contention, can be divided into clusters of data conflict-free transactions and a small set of residuals.

In this paper, we introduce a new concurrency control protocol called \system that leverages the above observation. \system executes transactions in batches, where each batch is partitioned into clusters of conflict-free transactions and a small set of residual transactions. The conflict-free clusters are executed in parallel without {\em any} concurrency control, followed by executing the residual cluster either serially or with concurrency control. We present a low-overhead algorithm that partitions a batch of transactions into clusters that do not have cross-cluster conflicts and a small residual cluster. We evaluate \system against the opportunistic concurrency control protocl and several variants of two-phase locking, where the latter is known to perform better than other concurrency protocols under high contention, and show that \system can improve transactional throughput by up to 2$\times$.
\end{abstract}

\maketitle

\section{Introduction}
\seclabel{introduction}

Online Transaction Processing (OLTP) Systems rely on the concurrency control protocol to ensure serializability of transactions executed concurrently. When two transactions executing in parallel try to access the same data item (e.g., a tuple, index entry, table, etc.), concurrency control protocol coordinates their accesses such that the final result is still serializable. Different protocols achieve this in different ways. Locking-based protocols such as two-phase locking (2PL), associate a lock with each data item and a transaction must acquire all locks (either in shared or exclusive mode) for data items it accesses before releasing any. Validation-based protocols such as optimistic concurrency control(OCC) ~\cite{occ}, optimistically execute a transaction with potentially stale or dirty (i.e. uncommitted) data and validate for serializability before commit. 

Validation-based protocols~\cite{hekaton, silo} are known to be well-suited for workloads with low {\em data contention} or {\em conflicts}, i.e., when data items are being accessed by transactions concurrently, with at least one access being a write. Since conflicting accesses are rare, it is unlikely that the value of a data item is updated by another transaction during its execution, and hence validation mostly succeeds. On the other hand, for workloads where data contention is high, locking-based concurrency control protocols are generally preferred as they pessimistically block other transactions that require access to the same data item instead of incurring the overhead of repeatedly aborting and restarting the transaction like in OCC. When the workload is known to be partitionable, partitioned concurrency control~\cite{hstore} is preferred as it eliminates the lock acquisition and release overhead for individual data items by replacing it with a single partition lock.

Recent empirical studies~\cite{dbx1000} have revealed that even 2PL-based protocols incur a heavy overhead in processing highly contended workloads due to lock thrashing (for ordered lock acquisition), high abort rates (for no-wait and wait-die protocols) or expensive deadlock-detection. Our main proposal in this paper is to eliminate concurrency control-induced overheads by intelligently scheduling these highly-contended transactions on cores such that the execution is serializable even \emph{without} any concurrency control. 

In this paper, we propose a new transaction processing scheme called \system that exploits data contention to improve performance in multi-core OLTP systems under high contention. The key insight behind \system is to recognize that most transactional workloads, even those with high data contention, can be partitioned into two portions: multiple {\em clusters} of transactions, where there are no data conflicts between any two clusters; and some {\em residuals} -- those that have data conflicts with atleast two other transactions belonging to different clusters.

As an example (to be elaborated in~\secref{architecture}), a workload that consists of \tpcc new order transactions can be divided into two portions: each set of transactions that orders from the same warehouse constitutes a cluster, while those that order from multiple warehouses constitute the residuals.

Since transactions in different clusters access disjoint sets of data items, they can be executed in parallel by assigning each cluster to a different core; each core can execute a given cluster {\em without any concurrency control}, and all of the executed transactions are guaranteed to commit (unless explicitly aborted). The residuals, on the other hand, can be executed serially either on a single core, again without any concurrency control, or across multiple cores with concurrency control applied. Our protocol aims to capture the ``best of both worlds'': partition the workload to identify as many clusters as possible as they can be executed without the overhead of running any concurrency control protocols, and minimize the number of residual transactions.

The idea of transaction partitioning is similar to partitioned databases~\cite{hstore}, where data items are split across different machines or cores to avoid simultaneous access of the same data items from multiple transactions. However, data partitioning needs to be done statically prior to executing any transactions, and migrating data across different machines during transaction execution is expensive. \system instead partitions transactions rather than data, and treats each batch of transaction as an opportunity to repartition, based on the access patterns that are inherent in the batch.

Furthermore, since data access is a property that can change over different workloads, \system is inspired by deterministic database systems~\cite{calvin} and executes transactions in {\em batches}. More precisely, \system collects transactions into batches; partitions the transactions into conflict-free clusters and residuals; and executes them as described above. The same process repeats with a new batch of transactions, where they are partitioned before execution.

Implementing \system raises a number of challenges. Clearly, a naive partitioning that classifies all transactions as residuals would fulfill the description above, although doing so will simply reduce to standard concurrency control-based execution and not incur any performance benefit. On the other hand, if the residual clusters are forced to be small, then number of conflict-free clusters produced might be lesser. As such, we identify the following  desiderata for our partitioning algorithm: 
\begin{myitemize}
    \item Minimize the number of residuals.
    \item Maximize the number and size of conflict-free clusters.
    \item Minimize the amount of time required to partition the transactions; time spent on partitioning takes away performance gain from executing without concurrency control.
\end{myitemize}

To address these challenges, \system comes with a novel algorithm to partition an incoming batch of transactions. It first represents the transaction batch as an {\em access graph}, which is a bipartite graph describing each transaction and the data items that are accessed. Partitioning then proceeds in $3$ steps: we first sample on the access graph to form the initial seed clusters, we then allocate the remaining transactions into the clusters. The resulting clusters are merged based on their sizes, and any leftover transactions are collected as the residuals. The final clusters are then stored in a worklist, with the cores executing them in parallel before proceeding to execute the residuals afterwards. Our prototype implementation has shown that the \system protocol can improve transaction throughput by up to 2$\times$, as compared to traditional protocols such as two-phase locking for high-contention workloads.

In summary, we make the following contributions:
\begin{myitemize}
    \item We propose a novel execution scheme for high contention workloads that is based on partitioning a batch of transactions into many conflict-free and a residual cluster. We use this clustering towards executing most transactions in the batch without any form of concurrency control except a few in the residual cluster.  
    \item We design a new algorithm for transaction partitioning based on their access patterns. Our algorithm uses a combination of sampling techniques and parallel data structures to ensure effiency.
    \item We have implemented a prototype of the \system concurrency control protocol, and evaluated using two popular transaction benchmarks: \tpcc~\cite{tpcc} and YCSB~\cite{ycsb}. The experiments show that \system can substantially improve the performance of transactional systems under high-contention by partitioning the workloads into clusters and residuals. 
\end{myitemize}

The rest of this paper is organized as follows: We first provide an overview of \system in~\secref{architecture}. Then in~\secref{algorithm} we discuss our partitioning algorithm in detail. We present our evaluation of \system in~\secref{evaluation}, followed by discussion of related work in~\secref{related-work}.
\section{Overview}
\seclabel{architecture}

In \system, transactions are scheduled on cores based on their data-access pattern. \system collects and executes transactions in batches, and assumes that read-write set of a transaction can be obtained statically. In scenarios where that is not possible, one can use a two-stage execution strategy similar to deterministic databases~\cite{calvin}: first dynamically obtaining the complete read-write set using a reconnaissance query, followed by a conditional execution of the transaction.

\system employs a \emph{micro-batch architecture} to execute transactions. Incoming transactions are grouped together into batches, partitioned into clusters and residuals, and scheduled to be executed on multiple cores. Micro-batching allows \system to analyze conflicting data accesses and utilize them to intelligently partition the workload.

\begin{figure}
    \centering
    \includegraphics[width=0.75\linewidth]{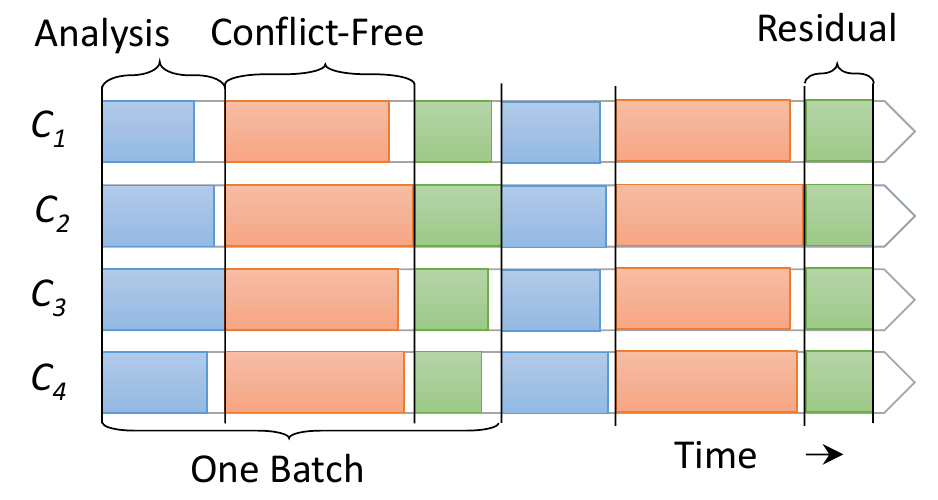}
    \caption{Execution Scheme of \system on $4$ cores}
    \figlabel{architecture}
\end{figure}
The execution scheme of \system is shown in \figref{architecture}.  A batch of transactions is executed in three phases: \analysis, \conflictfree and \residual phase. First, the batch of transactions is analyzed and partitioned into a number of \emph{conflict-free} clusters and a small \emph{residual}. Each conflict-free cluster is then executed without any concurrency control in parallel on all cores in the \conflictfree phase. After all clusters have finished execution, the residual transactions are executed on multiple cores with conventional concurrency control.\footnote{As mentioned in~\secref{introduction}, the residuals can be executed serially on a single core as well, although our experiments have shown that executing using multiple cores with concurrency control is a better strategy.} Once a batch is completed, \system repeats by analyzing the next batch.

We next give an overview of each of the three phases using an example workload.

\subsection{\analysis phase}
\begin{figure}
    \centering
    \includegraphics[width=0.9\linewidth]{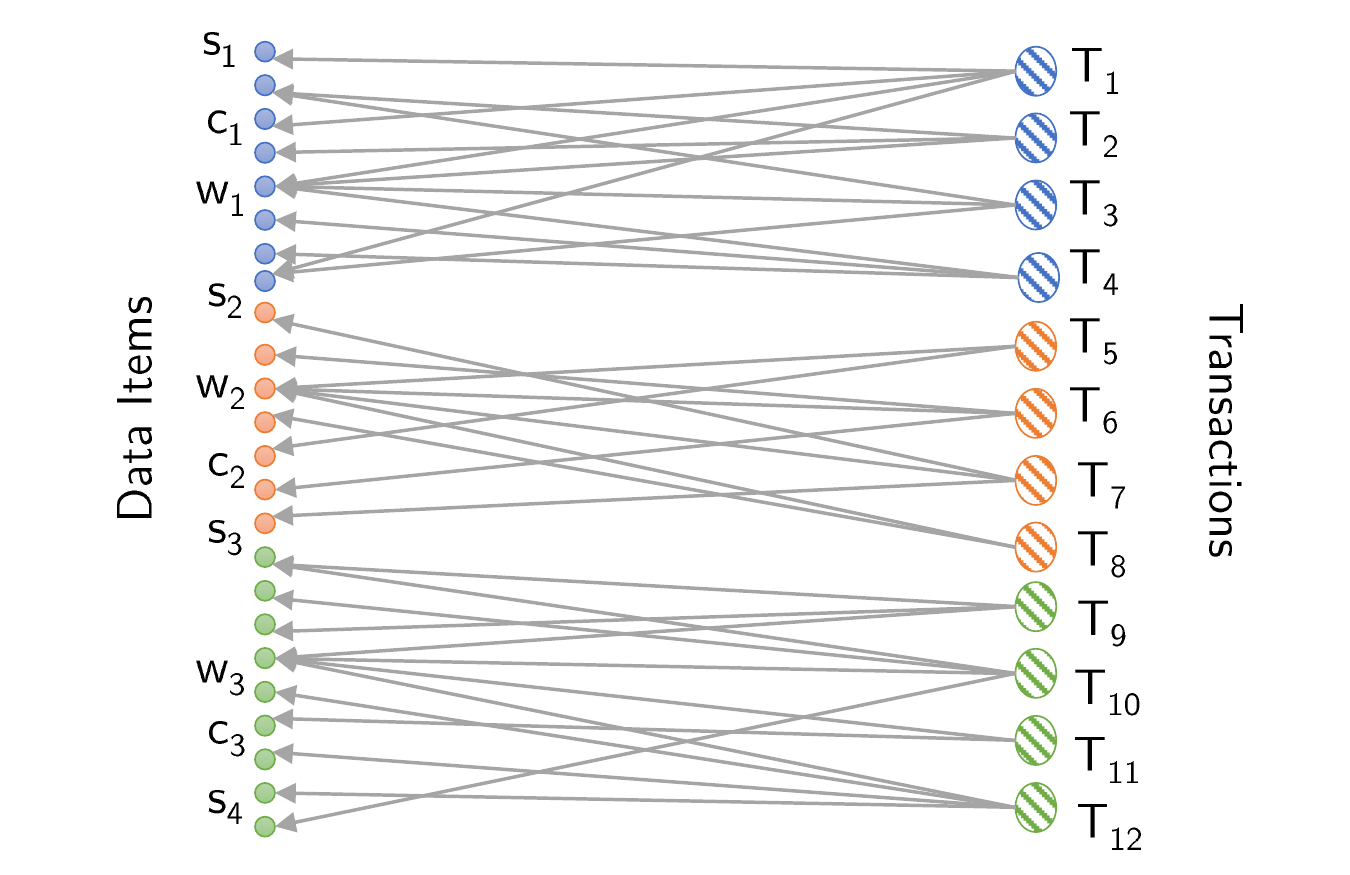}
    \caption{Access Graph of a \tpcc transactions}
    \figlabel{access-graph}
\end{figure}

\begin{figure*}
    \centering
    \begin{subfigure}[t]{0.45\textwidth}
    \includegraphics[width=\linewidth]{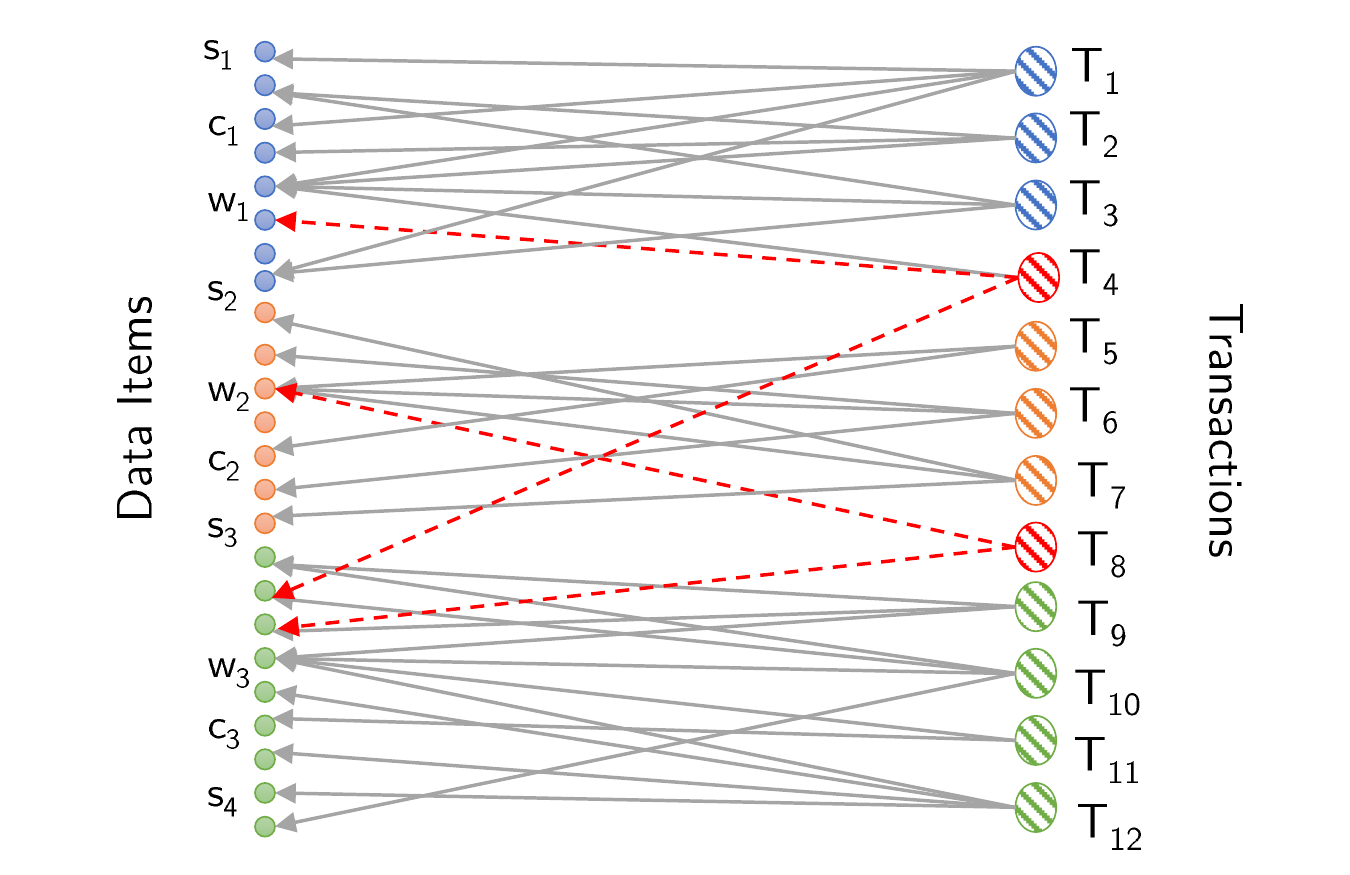}
    \end{subfigure}
    \begin{subfigure}[t]{0.45\textwidth}
    \includegraphics[width=\linewidth]{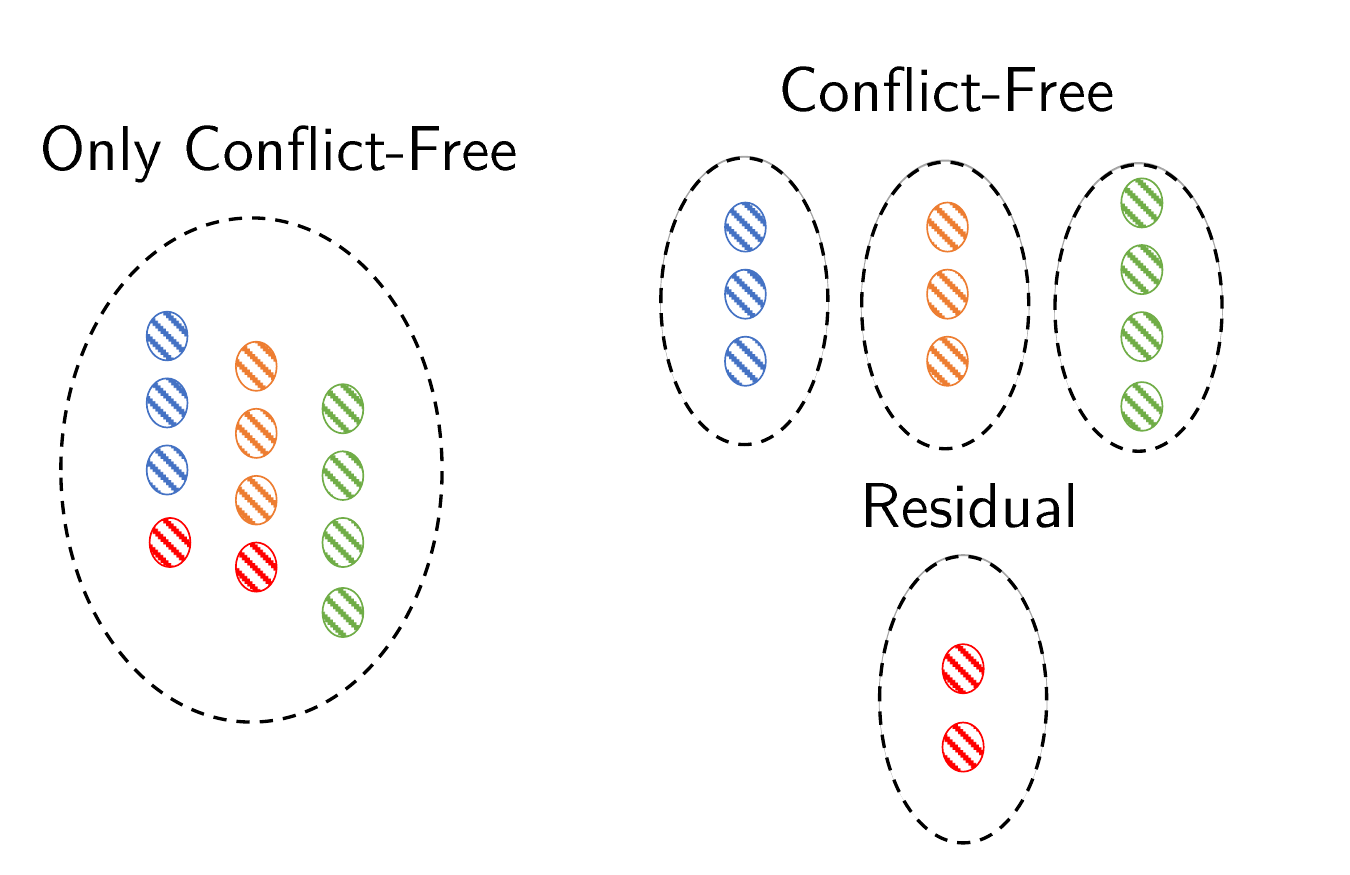}
    \end{subfigure}
    \caption{(a) Access Graph of \tpcc transactions (b) Optimal solutions for partitioning schemes}
    \figlabel{analysis}
\end{figure*}

The goal of the \analysis phase is to partition the batch of transactions into clusters such that any two transactions from two different clusters are conflict-free. We explain the details next.

To partition a batch of transactions, we first represent them using a {\em data access graph}. A data access graph is an undirected bipartite graph $\mathcal{A} = (\mathcal{B} \cup \mathcal{O}, \mathcal{E})$, where $\mathcal{B}$ is the set of transactions in the batch, $\mathcal{O}$ is the set of all data items (e.g., tuples or tables) accessed by transactions in $\mathcal{B}$, and the edges $\mathcal{E}$ contain all pairs $(T,d)$ where transaction $T$ accesses data item $d$.    Two transactions $T, T'$ are said to be \emph{in conflict} if they access a common data item and at least one of them is a write.

For example, \figref{access-graph} depicts the access graph of a batch of transactions from \tpcc benchmark. A \texttt{new-order} transaction simulates a customer order for approximately ten different items. The example in~\figref{access-graph} contains three different warehouses $w_1, w_2$ and $w_3$. Each warehouse maintains stock for a number of different items in the catalog.

As shown in the figure, transactions $\set{T_1, \ldots, T_{12}}$ access data items from different tables in the \tpcc database. $T_1$, for example, writes to the \texttt{warehouse} tuple $w_1$ and a few other tuples from other tables such as \texttt{district} and \texttt{stock} that belong to $w_1$ as well. Transactions $T_1, T_2$ are in conflict because they both access $w_1$; whereas transactions $T_1, T_5$ are not. The batch shown in \figref{access-graph}, is said to be \emph{partitionable} as groups of transactions access disjoint sets of data items. It can be partitioned into three clusters that do not conflict with each other, and the clusters can be executed in parallel with each one scheduled on a different core.

However, real workloads contain \emph{outliers} that access data items from multiple clusters. Consider the example shown in~\figref{analysis}(a), again of \tpcc {\tt new-order} transactions. Here, transactions $T_4$ and $T_8$ order items from multiple warehouses, resulting in a conflict with $T_2$ and $T_{10}$ respectively. There are two ways to execute $T_4$ and $T_8$: either merge the two clusters that $T_2$ and $T_{10}$ belong to and assign the resulting cluster to be executed on a single core, or move $T_4$ and $T_8$ into a separate cluster to be executed afterwards. As the former might result in a single large cluster that takes significant amount of time to execute, we take the latter approach where we consider $T_4$ and $T_8$ as residuals.
This results in the remaining batch partitioned into three conflict-free clusters along with the residuals, as shown in \figref{analysis}(b).

A \emph{clustering} is a partition of transactions $\mathcal{B}$ into $k+1$ sets $\set{C_1, C_2, \ldots, C_k, R}$ such that, for any $i \neq j$ and any transaction $T \in C_i, T' \in C_j$, $T$, $T'$ are not in conflict. Notice that no requirement is placed on the residuals $R$.  The data access graph does not distinguish between read and write access, because \system considers only data items for which there is at least one write by a transaction. Consequently, if any two transactions that access the same common item $d$ are placed in two distinct clusters, then at least one of them will have a write conflict with some other transaction, hence we do not need to consider the type of access to the data items.

During the \conflictfree phase, each cluster $C_i$ is executed on one dedicated core without any concurrency control between cores. After all clusters have finished, then, during the \residual phase, the residual transactions are executed, still concurrently, but with conventional concurrency control applied.  Ideally, we want $k$ to be at least as large as the number of cores to exploit parallelism at maximum during the \conflictfree phase, and we want $R$ to be empty or as small as possible to reduce the cost of the \residual phase.  To get an intuition about the tradeoffs, we describe two naive clusterings.  The first is the {\em fallback clustering}, where we place all transactions in $R$ and set $k=0$; this corresponds to running the entire batch using a conventional concurrency control mechanism.  The second is the {\em sequential clustering}, where we place all transactions in $C_1$, and set $k=1$ and $R=\emptyset$; this corresponds to executing all transactions sequentially, on a single core. As one can imagine, neither of these would result in any significant performance improvement. Hence in practice we constrain $k$ to be at least as large as the number of cores, and $R$ to be no larger than some small fraction $\alpha$ of the transaction batch. 

In practice, a good clustering exists for most transaction workloads, except for the extreme cases.
In one extreme, when all transactions in the batch access a very small number of highly contentious data items, then no good clustering exists besides fallback and sequential, and our system simply resorts to fallback clustering.  
Once the data contention decreases, i.e., the number of contentious data items $k$ increases to at least the number of cores, then a good clustering exists, and it centers around the contentious items.  When the contention further decreases to the other extreme where all transactions access different data items, then any partitioning of the transactions into $k$ sets of roughly equal size would be the best clustering.  Thus, we expect good clusterings to exist in all but most extreme workloads, and we validate this in ~\secref{evaluation}. The challenge is to find such a clustering very efficiently; we describe our clustering algorithm in~\secref{algorithm}.

\subsection{\conflictfree Phase}
After partitioning the incoming workload into conflict-free clusters, \system then schedules them to be executed in parallel on multiple \emph{execution threads}. Each execution thread obtains a conflict-free cluster and executes it to completion before moving to the next. Transactions belonging to the same cluster are executed serially one after another in the same execution thread.

Since the scheduling algorithm guarantees that there are no conflicts across transactions from different clusters, there is no need for concurrency control in this phase. As noted earlier, concurrency control is a significant overhead in transactional systems, especially for worloads that have frequent access to highly contended data. Hence removing it will significantly improve performance.

The degree of parallelism in this phase is determined by number of conflict-free clusters. Higher number of clusters result in them being executed in parallel, thereby reducing total time to execute all transactions in conflict-free clusters.

Once an execution thread has completed executing a cluster, it tries to obtain the next one. If there is none, it waits for all other threads that are processing conflict-free clusters before moving to the next phase. This is because residual transactions could conflict with transactions that are currently being executed by other \conflictfree phase threads without concurrency control. Hence, a skew in cluster sizes can cause a reduction in parallelism as threads that complete early cannot advance to next phase, although as our experiments show in~\secref{evaluation}, that is usually not the case.

\subsection{\residual phase}
As we saw in our example from~\figref{analysis}(a), \system identifies a few transactions to be outliers and considers them as residuals. These transactions conflict with transactions from more than one conflict-free cluster. We execute these residual transactions concurrently on all execution threads, but apply some form of concurrency control. Unlike the \conflictfree phase where the we guarantee conflict-freedom, transactions executed in the \residual phase require concurrency control to ensure serializability.

We could use any serializable concurrency control protocol in this phase. In \system, we use 2PL with \texttt{NO-WAIT} deadlock prevention policy as it has been shown to be highly scalable~\cite{dbx1000} with much less overhead compared to other protocols. Under this policy, when a transaction is aborted due to the concurrency control scheme, the execution thread retries it until successful commit or logical abort.

Once all the residual transactions have been executed, the same process is repeated with \system processes the next batch by running the \analysis phase.
\section{Transaction Partitioning Algorithm}
\seclabel{algorithm}
\begin{figure}
    \centering
    \includegraphics[scale=0.75]{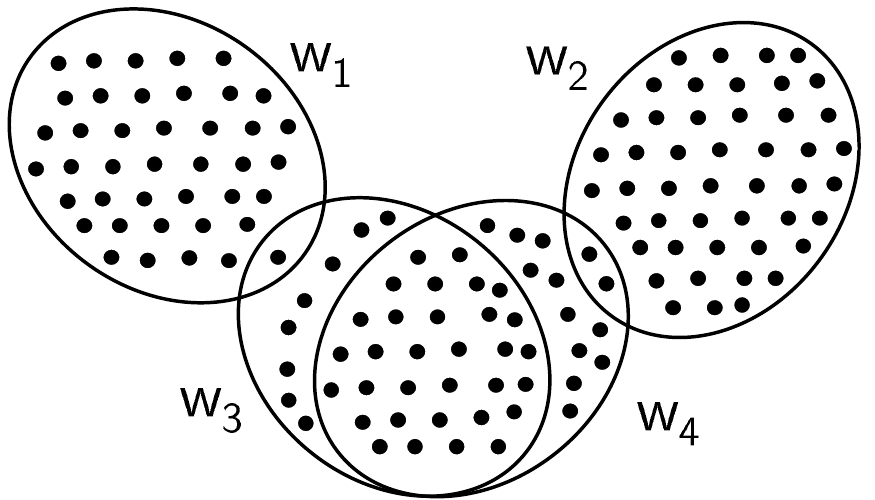}
    \caption{Example batch of \tpcc transactions}
    \figlabel{running-example}
\end{figure}

As mentioned, \system partitions a batch of transactions $\mathcal{B}$ into a set of $k$ conflict-free clusters $C_1, C_2, \ldots, C_k$, and a residual $R$ of size at most $\alpha |\mathcal{B}|$, with $\alpha$ being a configurable parameter. This partitioning problem can be modeled as graph partitioning on the data access graph $\mathcal{A}$ that corresponds to $\mathcal{B}$. Graph partitioning in general is NP-complete, and hence obtaining the optimal solution is exponential in the size of the batch.
Nevertheless, graph partitioning is a well researched area with many heuristic solutions. We review some of these solutions in \secref{related-work}. However, it is challenging to use an off-the-shelf solution to the problem at hand as most of the them do not meet the performance requirements in a low-latency transaction processing system. So, we developed a heuristic solution that exploits the contentious nature of each batch of transactions.

Our partitioning algorithm is divided into three stages: (1) \textsc{spot}, (2) \textsc{allocate} and (3) \textsc{merge}. In the \textsc{spot} stage, we identify highly contended data items from the batch using a randomized sampling strategy. Each of those data items are allocated to a different cluster. Transactions in the batch are then allotted to one of these clusters in the \textsc{allocate} phase. Finally, in the \textsc{merge} phase, some of these clusters are merged together to form larger clusters when a significant number of transactions in the batch co-access data items from multiple clusters.

We use \figref{running-example} derived from the \tpcc benchmark as an illustrative example. In the figure, a \texttt{new-order} transaction (black dots) shown inside a warehouse $w$ (circles) orders items only from warehouse $w$; and those that order from multiple warehouses are shown at their intersections. As shown in the figure, in the given batch the majority of transactions only orders locally from warehouses $w_1$ and $w_2$, while many transactions involving $w_3$ and $w_4$ order from multiple warehouses.

Before running the three stages of our partitioning algorithm, we first perform simple pre-processing on the transactions. During pre-processing step, \system receives incoming transactions, stores them in the current batch, and computes the set of data items that are accessed in a write mode by at least one transaction.  Data items that are read-only in the entire batch are ignored for partitioning purposes.
For example, \texttt{items} is a dimension table in the \tpcc benchmark that is mostly read and rarely updated; as a consequence many \texttt{items} elements in batch are ignored by our algorithm. In the rest of the algorithm, we consider only those data items $D_T$ in $T$ that are written to by at least one transaction in $\mathcal{B}$.

\subsection{Spot Stage}
\sseclabel{spot}
During the spot stage we create initial seeds for the $k$ clusters by randomly choosing $k$ mutually non-conflicting transactions. The pseudo-code for this stage is shown in \algref{spot-pseudo-code}.
Initially, all data items and transactions in the access graph $\mathcal{A}$ are unallocated to any cluster.
We begin by picking a transaction $T$ from $\mathcal{B}$ uniformly at random.
If none of the data items accessed by $T$, denoted $D_T$, is allocated to any cluster, then we create a new cluster $C_i$ and allot each $d \in D_T$ to cluster $C_i$.
If any of the data items is already allocated to a cluster, we reject $T$ for the next sample.
We repeat this randomized sampling of transactions for a constant $c$ number of times, where $c$ is some small factor times the number of cores.  When all transactions in the batch access a single highly contended data item $d$, for example, the initial pick will create cluster $C_1$ and allot $d$ to $C_1$. All future samples are now rejected as they access $d$. In such a case, we revert back to fallback clustering and execute $C_1$ sequentially.

The goal of the spot stage is to quickly identify highly contentious data items in the workload, as each such item should form their own cluster. To get some intuition of the working of the spot stage, suppose there are $k$ cores and the workload happens to have $k$ ``hot spot'' data items $d$, meaning data items that are each accessed by a fraction $1/k$ of all transactions in the batch.  An ideal clustering should place each hot spot in a different cluster $C_i$.  We observe that, in this case, with high probability, each of the hot spots is accessed by one of the transactions chosen by the spot stage as initial seeds. Indeed, every hot spot data item $d$ is picked with probability at least $1-(1-1/k)^c\approx 1$, because during each of the $c$ iterations, the probability that we chose a transaction that does not access $d$ is $1-1/k$ and, assuming $c \gg k$, we have $(1-1/k)^c \approx 0$.  This means that, with high probability, the spot stage will form cluster seeds that are centered precisely around those few hot spot data items.
By the same reasoning, if two hot spot data items are frequently accessed together, then with high probability the cluster that contains one will also contain the other.

\begin{algorithm}
\SetKwProg{Fn}{Function}{}{end}
\SetKwProg{Proc}{Procedure}{}{end}
\SetKwFunction{Cluster}{Cluster}
\SetKwFunction{SpotStage}{SpotStage}
\SetKwFunction{AllocateTxns}{AllocateTxns}
\SetKwFunction{AllocateSkippedTxns}{AllocateSkippedTxns}
\SetKwFunction{MergeClusters}{MergeClusters}

\Fn{\SpotStage{$\mathcal{B}$}}{
$\forall d:$ d.Cluster = NULL\;
R := $\{\}$ \tcp{Residual Cluster}
k := 1\;
\Repeat{$c$ times}
{
    Pick a random transaction $T$ from $\mathcal{B}$\;
    \If{$\forall d \in D_T:$ d.Cluster = NULL}
    {
        Create a new cluster $C_k$\;
        Add $T$ to $C_k$\;
        \ForEach{$d \in T$}
        {
            d.Cluster = $C_k$\;
        }
        k++\;
    }
}
\Return $(C_1, C_2, ..., C_k)$\;
}
\caption{Pseudo-Code for the Spot Stage}
\alglabel{spot-pseudo-code}
\end{algorithm}
In our example, it is we pick one of $w_1$ or $w_2$-only transaction in one of the rounds with a high probability. So, any $w_1$ or $w_2$ transaction picked in the future is simply rejected as the corresponding \texttt{warehouse} tuple is already allotted a cluster. Similarly, a $w_3 - w_4$ transaction might be picked in one of the rounds resulting in three base clusters. At this stage, further sampling of transactions does not increase the number of base clusters as all other transactions will be rejected. In an alternate scenario, a $w_3$-only and a $w_4$-only transaction might be picked before any $w_3-w_4$ transaction due to the randomness of the event resulting in $4$ base clusters.

\subsection{Allocate Stage}
\sseclabel{allocate}
In this stage, we develop on the initial seed clusters created previously by allocating more transactions and data items accessed by them to these clusters.

In this stage, we allocate transactions in two rounds. Let the seed clusters be $\{C_1, C_2, ..., C_k\}$. In the first round, we scan through transactions in $\mathcal{B}$ and try to allot a previously unallocated transaction $T$ to one of these clusters or as a residual based on the following criteria (refer \algref{allocate-pseudo-code} for details):
\begin{myitemize}
\item If none of the allocated transactions access data items in $D_T$, then we leave $T$ unallocated.
\item If all of the data items in $D_T$ are allocated to a unique cluster $C$, then we allocate $T$ to $C$ as well and allocate all the other data items in $D_T$ to $C$.
\item When data items in $D_T$ are allocated to more than one cluster, we allot $T$ to residuals $R$.
\end{myitemize}

Let the \emph{distance} between two transactions $T$ and $T'$, denoted $\eta(T, T')$, be $\frac{1}{2}$ the length of shortest path between them in $\mathcal{A}$. For example, distance between $T_1$ and $T_2$ in \figref{analysis}(a) is 1 due to $T_1 - w_1 - T_2$.  Distance between $T$ and cluster $C$ is the shortest distance between any transaction in $C$ and $T$.

At the end of first round, all transactions that are at a distance $1$ from initial seed clusters are allocated to one of $C_1, C_2, ..., C_k$ or $R$. If $\kappa$ is the maximum distance between two transactions in the same connected component, then repeating the above allocation round for $\kappa$ times will eventually allocates all transactions. However, we observe that in practice, $\kappa$ is close to $1$ for high contention workloads. Hence, in many cases we only need to run the above allocation mechanism once.

Next, we handle the remaining transactions in $\mathcal{B}$ are left unallocated after the above process has taken place. To allocate them, we run a second round of allocation, but with a slight modification. Instead of skipping a transaction when $D_T$ is unallocated, we allot it to one of the $C_1, ... C_k$ clusters randomly. Transactions that were at a distance of $2$ from initial seed clusters are now at a distance of $1$ as new transactions have been allocated to them in the first round. So, some unallocated transactions will now have allocated data items. These are processed as in the first round: allocate to $C$ if it is the unique cluster for data items in $D_T$ and to $R$ if data items in $D_T$ are allocated to more than one cluster.

At the end of the \textsc{allocate} stage, we have a set of clusters $\{C_1, ..., C_m\}$ (where $m \geq k$) and residual transactions $R$ such that all transactions in a cluster $C_i$ access data items only in $C_i$, and the transactions in $R$ access data items belonging to more than one of the $m$ clusters.

\begin{algorithm}
\SetKwProg{Fn}{Function}{}{end}
\SetKwProg{Proc}{Procedure}{}{end}
\SetKwFunction{Cluster}{Cluster}
\SetKwFunction{AllocateStage}{AllocateStage}
\SetKwFunction{AllocateTxns}{AllocateTxns}
\SetKwFunction{AllocateSkippedTxns}{AllocateSkippedTxns}
\SetKwFunction{MergeClusters}{MergeClusters}

\Fn{\AllocateStage{$\mathcal{B}$}}{

\ForEach{$T \in B$}
{
    \uIf{$\forall d \in D_T:$ d.Cluster = NULL}
    {
       skip\;
    }
    \uElseIf{$\forall d \in D_T:$ d.Cluster = NULL or unique $C$}
    {
        Add $T$ to $C$\;
        \ForEach{$d \in  D_T$ and d.Cluster = NULL}{
            d.Cluster = $C$\;
        }
    }
    \Else
    {
        Add $T$ to $R$ \;
    }
}

\ForEach{Unallocated $T \in B$}
{
    \uIf{$\forall d \in D_T:$ d.Cluster = NULL}
    {
       Pick a random $C$ from $C_1, C_2, ..., C_k$\;
       Add $T$ to $C$\;
       \ForEach{$d \in  D_T$ and d.Cluster = NULL}{
            d.Cluster = $C$\;
        }
    }
    \uElseIf{$\forall d \in D_T:$ d.Cluster = NULL or unique $C$}
    {
        Add $T$ to $C$\;
        \ForEach{$d \in  D_T$ and d.Cluster = NULL}{
            d.Cluster = $C$\;
        }
    }
    \Else
    {
        Add $T$ to $R$ \;
    }
}

\Return $(C_1, ..., C_k, R)$\;
}
\caption{Pseudo-Code for the Allocate Stage}
\alglabel{allocate-pseudo-code}
\end{algorithm}

In our \tpcc example, if the \textsc{spot} stage had resulted in $3$ base clusters one each for $w_1, w_2$ and $w_3-w_4$, then most transactions in $\mathcal{B}$ will be allocatted to one of the $3$ clusters in the \textsc{allocate} stage. A small number of transactions that are $w_1-w_3$ or $w_2-w_4$, however, will not be allocatted to either of these clusters and will be added to the residuals. If the \textsc{spot} stage produced $w_1, w_2, w_3$ and $w_4$ as base clusters, most of $w_1$ and $w_2$ transactions from $\mathcal{B}$ will be allocatted to its clusters.  However, none of the $w_3-w_4$ transactions can be added to any of the clusters and hence will be added to the residual cluster. We further process the resulting clusters from this stage to reduce the size of the residual cluster. Our example does not execute the second round as all transactions are aready allocated during the first round.

\subsection{Merge Stage}
Depending on the nature of base clusters created in \textsc{spot} stage, the number of residual transactions  that remain at the end of the \textsc{allocate} stage could be large. During \textsc{merge}, we merge some of these clusters to improve the quality of the clusters and to reduce the size of residual cluster. When two clusters $C_i$ and $C_j$ are merged to form a new cluster $C_k$, transactions in $R$ that access data items only from $C_i$ and $C_j$ can now be allocated to $C_k$ using the allocation criteria mentioned above. 

While merging reduces the number of residual transactions, excessive merging of clusters could result in forming one large cluster which reduces parallelism in \conflictfree phase. Hence, we merge clusters until size of the residual cluster is smaller than the bound specified by the parameter $\alpha$, i.e. $|R| \leq \alpha |\mathcal{B}|$ using the scheme detailed in \algref{merge-pseudo-code}.  $\alpha$ serves as a parameter that chooses between executing transactions on multiple cores with concurrency control (if $\alpha$ is small) versus on fewer cores but with no conflicts and without concurrency control (otherwise). Empirically, we found $\alpha = 0.2$ to be appropriate in our experiments.

Let $N(C_i, C_j)$ denote the number of transactions in $R$ that access data items in $C_i$ and $C_j$. Note that the transactions that are accounted for in $N(C_i, C_j)$ can access data items from clusters other than $C_i$ and $C_j$ as well. If the two clusters $C_i$ and $C_j$ are separated, then all of the $N(C_i, C_j)$ transactions will be marked as residuals. So, we merge cluster pairs $C_i, C_j$ using the following criterion:
\begin{equation*}
    N(C_i, C_j) > \alpha \times (|C_i| + |C_j| + N(C_i, C_j))
\end{equation*}
Since, $|R| \leq \sum_{i \neq j} N(C_i, C_j)$, a merge scheme using the above criterion always results in the number of residuals being smaller than $\alpha |\mathcal{B}|$. Once all such clusters have been merged, transactions in the residual cluster are re-allocated to the new clusters when all data items accessed by a transaction belong one unique cluster. The resulting conflict-free clusters are then executed in parallel without any concurrency control, followed by the residuals with concurrency control applied as discussed in \secref{architecture}.

\begin{algorithm}
\SetKwProg{Fn}{Function}{}{end}
\SetKwFunction{PivotStage}{PivotStage}
\SetKwFunction{MergeStage}{MergeStage}
\Fn{\MergeStage{$C_1, ..., C_k, R$}}
{
Clusters := $\{ C_1, C_2, ..., C_k \}$\;
\ForEach{$C_i, C_j : N(C_i, C_j) \geq \alpha * (|C_i| + |C_j| + N(C_i, C_j))$}
{
    Create new cluster $C$\;
    $C$ := $C_i \cup C_j$\;
    Remove $C_i, C_j$ from Clusters\;
    Add $C$ to Clusters\;
    \ForEach{$d:$ d.Cluster = $C_i$ or $C_j$}
    {
        d.Cluster = $C$\;
    }
}
\BlankLine
\ForEach{$T \in R$}
{
    \If{$\forall d \in D_T: d.Cluster = C$}
    {
        Add $T$ to $C$\;
    }
}

\Return{(Clusters, $R$)}\;
}
\caption{Pseudo-code for the Merge Stage}
\alglabel{merge-pseudo-code}
\end{algorithm}
In our example, when the base clusters are $w_1, w_2, w_3-w_4$, the number of transactions that are allotted as residuals are small, and hence there is no merging of clusters needed. However, if the clusters are $w_1, w_2, w_3, w_4$, then size of the residual cluster is large and clusters $w_3$ and $w_4$ are merged. None of the other clusters are merged together as they do not satisfy the merge criterion. The final clusters are then $w_1, w_2$ and $w_3-w_4$ with a small amount of transactions ($w_1-w_3$ and $w_2-w_4$) in the residual cluster. In this example, our algorithm has essentially identified $3$ conflict-free clusters that can now executed without any concurrency control, where all transactions in these clusters access hot data items.
\section{Evaluation}
\seclabel{evaluation}

We have implemented a prototype of \system and evaluated the following aspects of \system:
\begin{myitemize}
\item We compare the performance of \system with variants of the two-phase locking protocol for high contention workloads. The results show that \system achieves up to $2\times$ better throughput both on YCSB and \tpcc benchmark.
\item We study the impact of the number of ``hot'' records on performance by varying the number of partitions in the YCSB mixture and number of warehouses in \tpcc. We show that \system is able to improve its performance as the number of hot items that are mostly independently accessed increases.
\item We characterize the scalability of \system along with other protocols by varying the number of threads in the system for a highly contended YCSB and \tpcc workload and \system outperforms traditional protocols by $2\times$ in terms of throughput.
\item We evaluate the impact of contention by varying the zipfian constant of the YCSB workload. We observe that while other 2PL protocols perform better at lower contention workload, \system outperforms them by up to $4\times$ in throughput when the contention is higher.
\end{myitemize}

\subsection{Implementation}
We have implemented a prototype of \system that schedules transactions based on the algorithm described above. \system, at its core, is a multi-threaded transaction manager that interacts with the storage layer of the database using the standard \texttt{get-put} interface. A table is implemented as a collection of in-memory pages that contain sequentially organized records. Some tables are indexed using a hashmap that maps primary keys to records. We implement the index using libcuckoo ~\cite{libcuckoo} library, a thread-safe fast concurrent hashmap. Records contain additional meta-data required to do scheduling and concurrency control. We chose to co-locate this with the record to avoid overheads due to multiple hash lookups and focus primarily on concurrency control.

As discussed earlier, \system groups transactions together into batches and process them in a three phases.

\paragraph{Partitioning} We prioritize minimizing the cost of \analysis phase over the optimality of our scheduling algorithm. Threads synchronize after each of the three stages of the algorithm. First, the \spot stage is executed using a single thread, followed by \allocate in parallel. The batch is partitioned into equal chunks and a thread allocates transactions in its chunk to the base clusters created in \spot phase. Each record has a \texttt{clusterId} as part of its meta-data and is updated atomically using an atomic \emph{compare-and-swap} operation. A transaction is allotted to a cluster only when all atomic operations on a transaction succeeds.

This is followed by the \merge stage that is carried out by a single thread. The cluster pair counts used in \merge stage are gathered during the \allocate phase using thread-local data structures and finally merged to obtain the global counts. Each cluster has a root, which initially points to itself. When merging two clusters, we modify the root of one cluster to point to another. To obtain the cluster to which a record or transaction belongs, we trace back to the root.
Finally, similar to the \allocate phase, the residual transactions are re-allocated to clusters in parallel.

\paragraph{Execution} The analysis phase produces a set of conflict-free clusters and the residuals. The conflict-free clusters are stored in a multi-producer-multi-consumer concurrent queue, called the \emph{worklist}. Each thread dequeues a cluster from the worklist, executes it completely without any concurrency control and obtains the next. Threads wait until all other threads have completed executing the conflict-free clusters.

Once the conflict-free clusters are executed, threads then execute the residual transactions. The residuals are stored in a shared concurrent queue among the threads. Threads dequeue a transaction, and execute it using the two-phase locking concurrency control under the \texttt{NoWait} policy (i.e., immediately aborts the transaction if it failes to grab a lock). \system moves to the \analysis phase of next batch once the \residual phase is completed. Technically, the threads can start analyzing the next batch while \residual phase of previous batch is in-progress. However, we did not implement this optimization to simplify the analysis and interpretability of results.  

\subsection{Experimental Setup}
We run all our experiments on a multi-socket Intel (R) Xeon(R) CPU E7-4890 v2 @ 2.80GHz with 2TB Memory. Each socket has 15 physical and 30 logical cores. All our experiments are limited to cores on a single socket.

We implemented \system and our baselines in C++. 
In all our experiments we set the batch size to be $100$K transactions resulting in a latency of atmost $200$ms. Note that this is lower than recommended client response time of $500$ms for the \tpcc benchmark~\cite{tpcc}, and it did not result in any significant difference in the results.

\subsection{Workloads}
All our experiments use the following workloads:
\begin{myitemize}
\item \textbf{TPC-C}: We use the a subset of the standard \tpcc benchmark. We restrict our evaluation to a 50:50 mixture \texttt{New-Order} and \texttt{Payment} transactions. 
We pick these two transactions as they are short ones that can stress the overhead of using a concurrency control protocol.

All tables in \tpcc have a key dependency on the \texttt{Warehouse} table, except for the \texttt{Items} table. Hence most transactions will access at least one of the warehouse tuple during execution. Each warehouses contains $10$ districts, each district contains $10$K customers. The catalog lists $100$K items and each warehouse has a stock record for each item. Our evaluation adheres to the \tpcc standards specification regarding remote accesses: $15$\% payment transactions are to remote warehouses and $1$\% of items are ordered from remote warehouses. Each \texttt{New-Order} transaction orders approximately $10$ items resulting in a total of $10$\% remote stock accesses. We do not, however, use secondary index to retrieve a customer record using last name and restrict to querying by the primary key \texttt{customer\_id} only.
\item \textbf{YCSB}: The YCSB workload is designed to stress the concurrency control further and help in various micro-benchmark experiments. YCSB transactions are a sequence of read-write requests on a single table. The table contains $10$M keys with a payload size of $128$~bytes and is queried using its primary key. Transactions are generated as belonging to specific partitions where the intra-partition key-access is determined by a zipfian distribution. The distribution can be controlled using the \emph{zipfian constant}, denoted using $\theta$. The higher the value of $\theta$, the higher the frequency of accessing the hotter keys in the distribution. Each transaction consists of $20$ accesses with a $50$\% probability of reads vs. writes. In our experiments, we control the number of hot records by varying number of partitions.
\end{myitemize}

\subsection{Baselines}
We compare our \system prototype with variants of the two-phase locking (2PL) protocol. Several experimental studies~\cite{dbx1000} have shown that 2PL strategies outperform other validation or multi-version based protocols for highly contented workloads. Below are the implementation specifics of our baselines:
\begin{myitemize}
    \item \textbf{NoWait}: NoWait is a variant of the 2PL protocol where a transaction acquires shared or exclusive locks depending on the access type during execution, and releases them only upon commit or abort. If a transaction is unable to acquire a lock, it is aborted immediately (without waiting) to prevent deadlocks. We use as many locks as the number of records in the database, each co-located with the record to avoid overheads due to a centralized lock table.

    \item \textbf{WaitDie}: WaitDie is another variant of 2PL that uses a combination of abort and wait decisions based on timestamps to prevent deadlocks. A transaction requesting for a shared or exclusive lock waits in the queue corresponding to the lock only when its timestamp is smaller than all the current owners of the data item. Since transactions in the waiting queue always have decreasing timestamps, there is no deadlock possible. We use a hardware counter-based scalable timestamp generation technique proposed in prior work~\cite{dbx1000}.

    \item \textbf{LockOrdered}: This is a deadlock-free variant of 2PL. Before execution, a transaction acquires shared or exclusive locks on all data items it accesses in some pre-determined order to prevent deadlocks, and releases them after committing the transaction.

    \item \textbf{WaitsForGraph}: We use a graph, called the \emph{waits-for graph} to track dependencies between transactions waiting to acquire logical locks and their current owners. Each database thread maintains a local partition of the wait-for graph similar in prior work~\cite{dbx1000}. A transaction is added to the waits-for graph only when a lock is currently held in a conflicting mode by other transaction(s). A cycle in the dependency graph implies a deadlock and the recently added transaction is aborted.
\end{myitemize}

\subsection{Varying number of hot records}
\sseclabel{hot-records}
\begin{figure*}
    \centering
    \begin{subfigure}[t]{0.45\textwidth}
    \includegraphics[width=\linewidth]{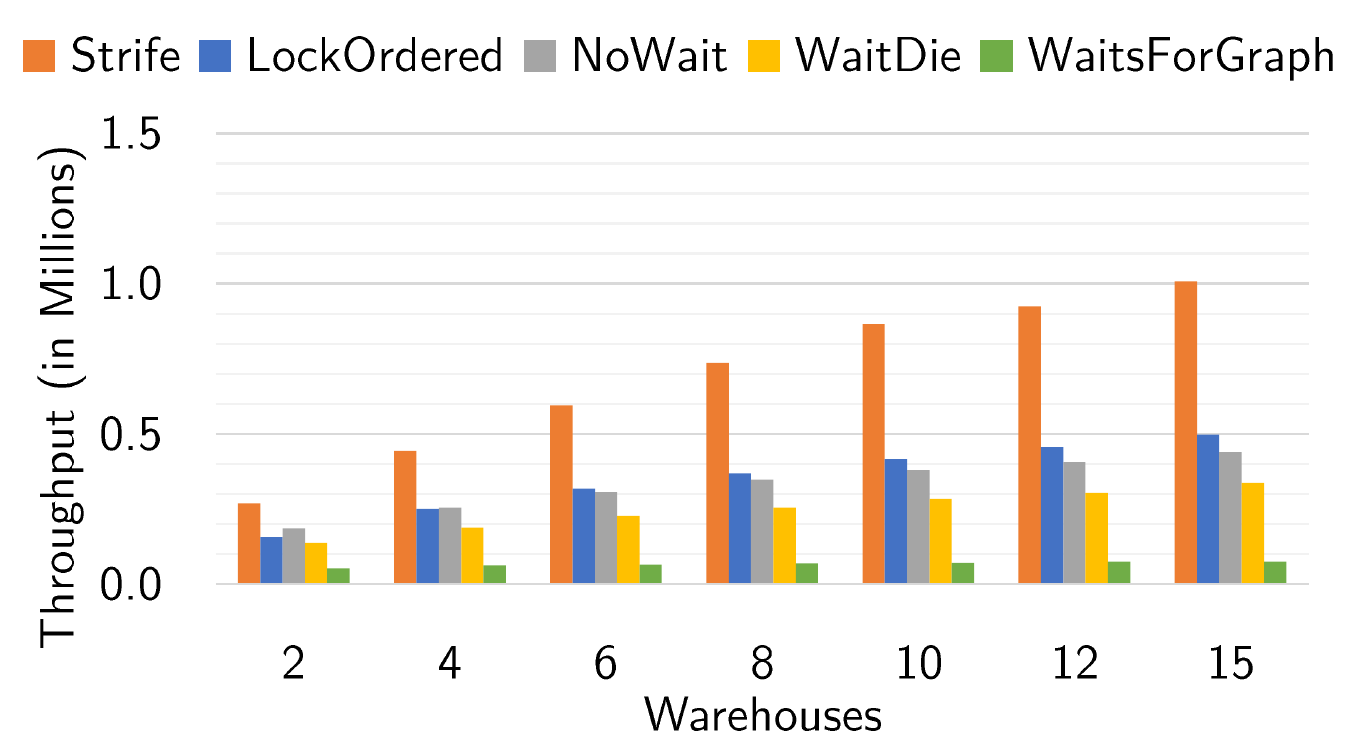}
    \figlabel{tpcc-contention-main}
    \end{subfigure}
    \hspace{0.5cm}
    \begin{subfigure}[t]{0.45\textwidth}
    \includegraphics[width=\linewidth]{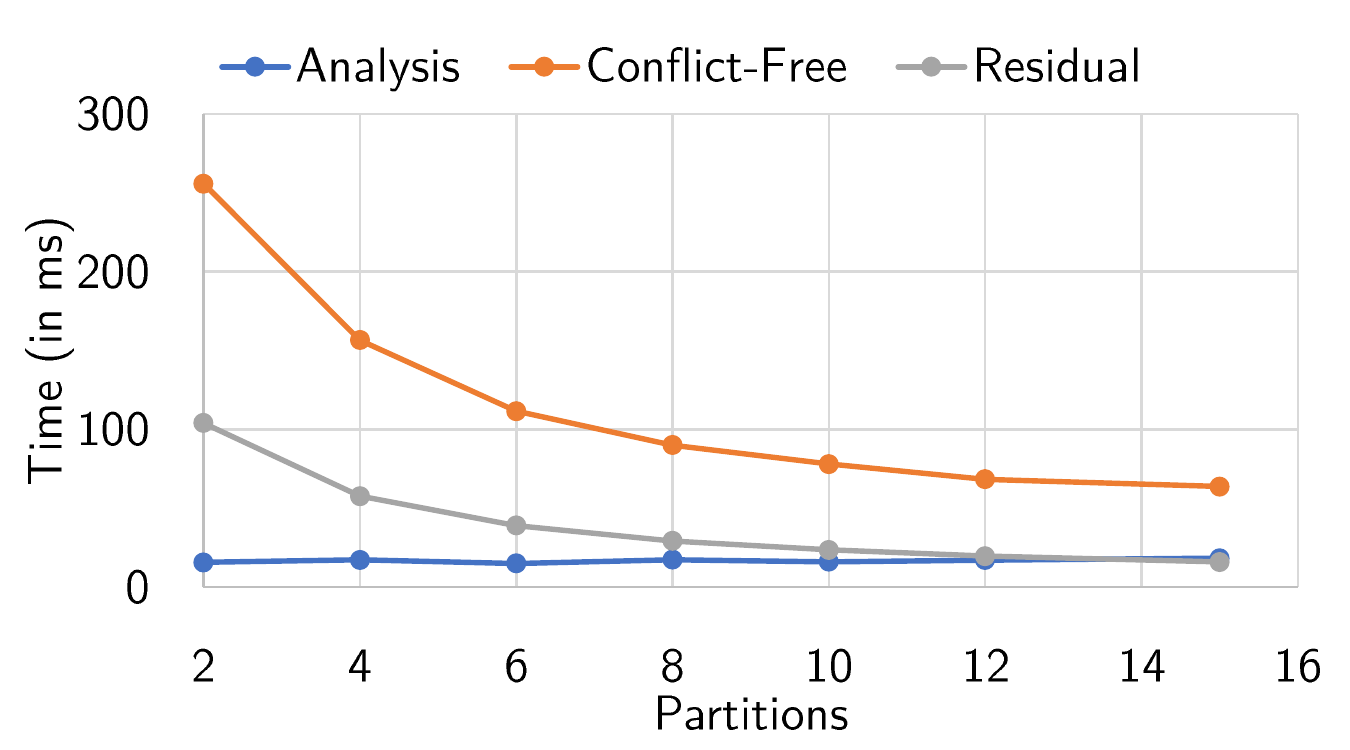}
    \figlabel{tpcc-contention-pipeline}
    \end{subfigure}
    \caption{Performance of \tpcc on $15$ cores: (a)Throughput vs. Number of Warehouses (b) Runtime Breakdown}
    \figlabel{tpcc-contention}
\end{figure*}

\begin{figure*}
    \centering
    \begin{subfigure}[t]{0.45\textwidth}
    \includegraphics[width=\linewidth]{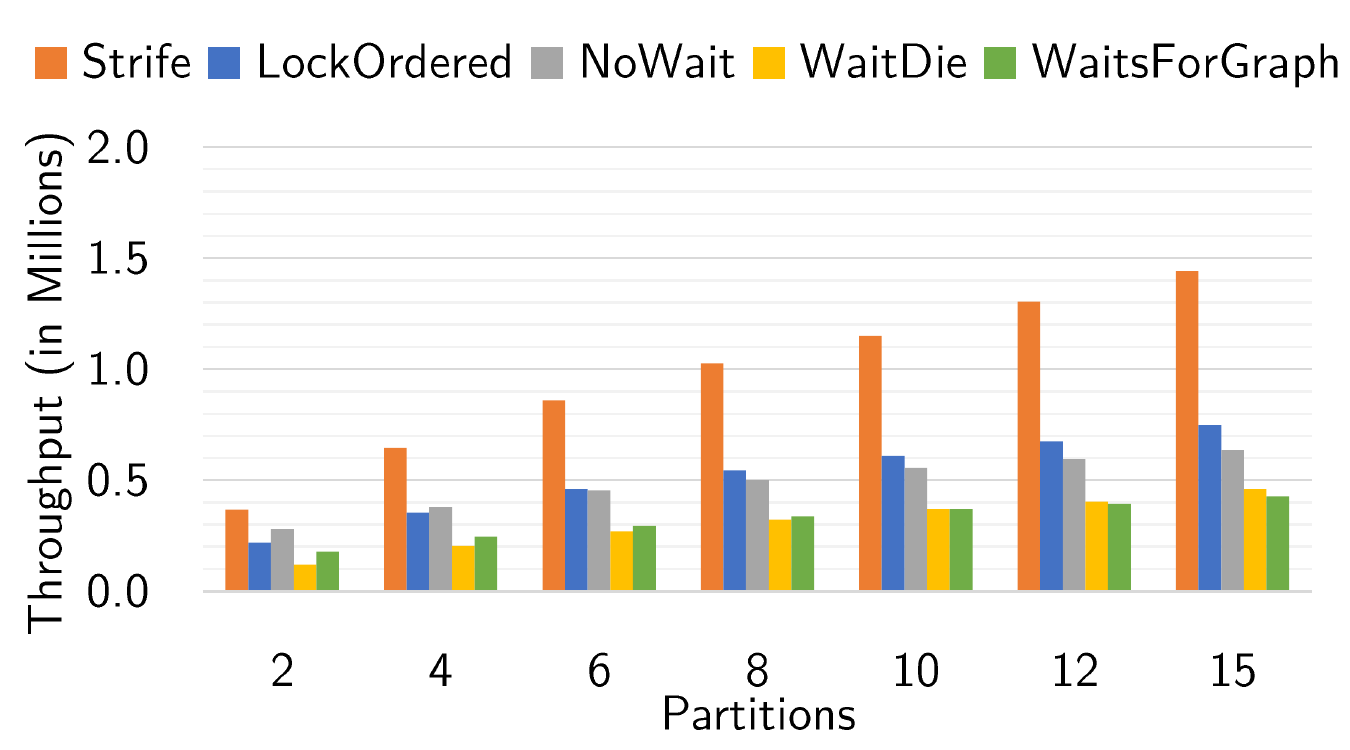}
    \figlabel{ycsb-contention-main}
    \end{subfigure}
    \hspace{0.5cm}
    \begin{subfigure}[t]{0.45\textwidth}
    \includegraphics[width=\linewidth]{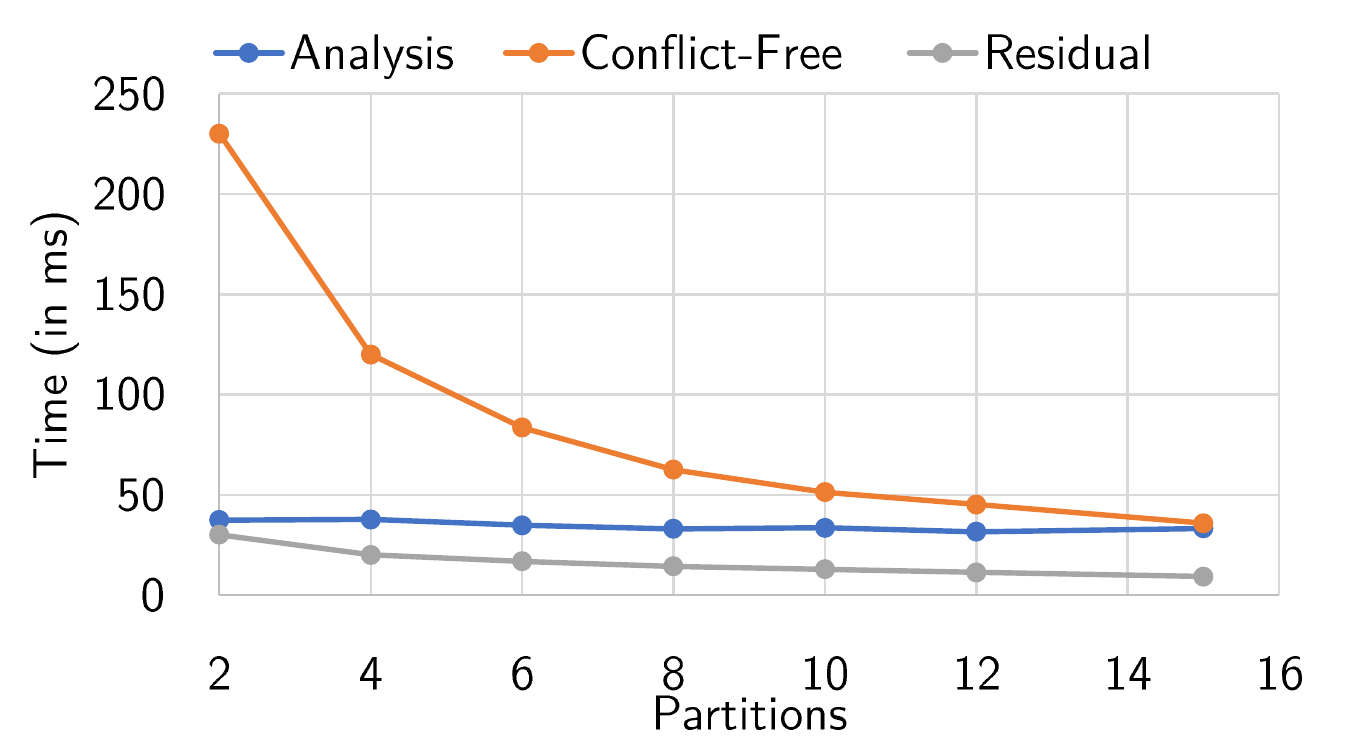}
    \figlabel{ycsb-contention-pipeline}
    \end{subfigure}
    \caption{Performance of YCSB on $15$ cores: (a)Throughput vs. Number of Warehouses (b) Runtime Breakdown}
    \figlabel{ycsb-contention}
\end{figure*}

We first analyze the performance of \system and compare it with our baseline concurrency control protocols under high contention by varying number of hot records.

Contention in \tpcc workload can be controlled by varying the number of records in the warehouse table as all transactions access (i.e., read or write) the warehouse tuple. 
\figref{tpcc-contention}(a) shows throughput in number of transactions committed per second vs. number of warehouses. 
As the number of warehouses increases from left to right, contention in the workload \emph{decreases} from left to right.
Payment transaction updates two contentious records: district and warehouse, while a new order reads warehouse and items tuples and updates district, customer, and other low contention items from stock table. 
In our experimental setup, we retry a transaction if aborts due to the concurrency control. 

The results of the experiment are shown in~\figref{tpcc-contention}(a) (for \tpcc) and~\figref{ycsb-contention}(a) (for YCSB) respectively. The results show that \system significantly outperforms all other protocols by up to $2\times$ in terms of throughput. When contention decreases, any concurrency control protocol is expected to improve in performance. Specifically, the number of warehouses in the workload determines the number of conflict-free clusters produced by the \analysis phase. When the number of warehouses (in \tpcc) or partitions (in YCSB) is greater than the number of available cores (15 in our experiments), the conflict-free clusters are executed in parallel without any concurrency control. However, other protocols are unable to exploit this parallelism as well as \system because the workload still have significant number of conflicts within each warehouse. 

We now explain the results in detail. The LockOrdered protocol is based on ordered acquisition of logical locks on records. A thread spin-waits until a requested lock is granted. When the number of warehouses is $2$, most threads are blocked except for the two that current have ownership of locks on the warehouses. So, the performance of LockOrdered is poor when the number of warehouses is small. However, as we increase the warehouse from $2$ to $15$, the chance that a thread $T$ is blocked decreases by a factor of $\frac{2}{15}$, so the LockOrdered protocol is seen to recover the performance outperforming other 2PL variants. On the other hand, \system eliminates the locking overhead, and thus results in much better performance. 

NoWait and WaitDie protocols use aborts to avoid deadlocks. The advantage of NoWait over WaitDie is that the former has very little overhead as it only needs a single record-level mutex for concurrency control. Hence aborts are cheap and even repeated retries are not very expensive. The WaitDie protocol incurs additional overhead in the form of waiting queue for each record. Another reason for the poor performance of WaitDie is that when a transaction with timestamp $t$ gets aborted as it is younger than the current owner, it is also younger than the waiters and hence during retry, it is highly likely that it aborts again. We observe the abort rate of WaitDie is more than $50$\% in our experiments. 

The WaitsForGraph is more conservative regarding aborting a transaction. It aborts a transaction only when there is a cycle in the global waits-for graph. Even though the graph is maintained in a scalable thread-local fashion, deadlock detection acquires locks on the data structures of other threads and hence serves as a concurrency bottleneck. Note that in \tpcc the actual occurrence of deadlocks is rare and cycle detection is purely an overhead. 

\figref{tpcc-contention}(b) depicts the average time taken by each phase in \system for a batch of size $10$K transactions. The cost of analysis is almost constant as we vary the number of partitions. However, the residual phase time and hence the number of residual transactions drops steadily. This is because when a new order transaction $T$ belonging to warehouse $w$ orders an item from a remote warehouse $w'$, it accesses the corresponding stock $s$. It is considered an \emph{outlier} access only if there also exists a local transaction to $w'$ that accesses the same $s$. Otherwise $s$ will be part of $T$'s cluster and considered to be conflict-free. So, when the number of warehouses are small, there is a high probability that this happens and hence more residual transactions. We also observe that the time for \conflictfree phase decreases steadily as we increase the number of partitions. This further validates that \system exploits parallelism even in the high-contention workload.

Next, we perform a similar experiment on the YCSB workload by varying the number of partitions. We use a zipfian constant of $0.9$ to produce a highly contended workload. The main difference between \tpcc and YCSB workload is that all transactions access one of the highly contended warehouse tuple in \tpcc, thereby reducing the diameter of the access graph of the batch to $1$. Whereas in YCSB, transactions belonging to a partition $p$ need not all access a single contentious data. The zipfian distribution creates a set of hot items in each partition with varying degrees of ``hotness".

Finally, \figref{ycsb-contention}(a) shows the comparison of \system with the baselines. The observations are largely similar to \tpcc. As the number of partitions increases, the total amount of contention in the batch decreases. While this improves the performance of all protocols, \system still outperforms others with up to $2\times$ improvement in throughput, despite the fact that unlike \tpcc, almost $50$\% of time is spent in \analysis for the $15$ partitions. The \conflictfree phase steadily decreases as the batch can be executed in higher degrees of parallelism. We also note that even though the batch is completely partitionable into $15$ clusters by design, \analysis phase produces more than $15$ clusters resulting in some single partition transactions being labeled as residuals.

\subsection{Scalability}
\sseclabel{scalability}
In this section, we analyze the scalability of \system and baselines on a high contention workload. We set the number of warehouses to be $4$ in \tpcc and vary number of cores. 

\begin{figure*}
    \centering
    \begin{subfigure}[t]{0.45\textwidth}
    \includegraphics[width=\linewidth]{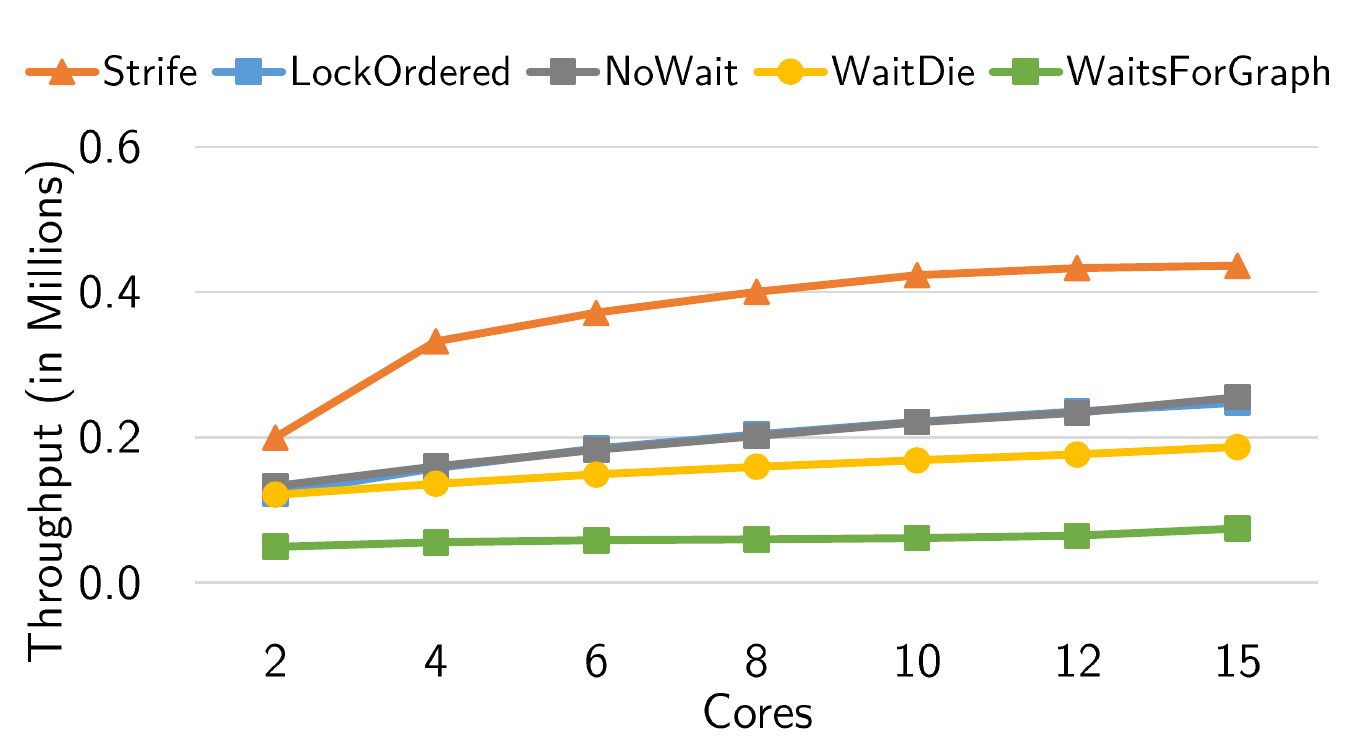}
    \figlabel{tpcc-scalability-main}
    \end{subfigure}
    \hspace{0.5cm}
    \begin{subfigure}[t]{0.45\textwidth}
    \includegraphics[width=\linewidth]{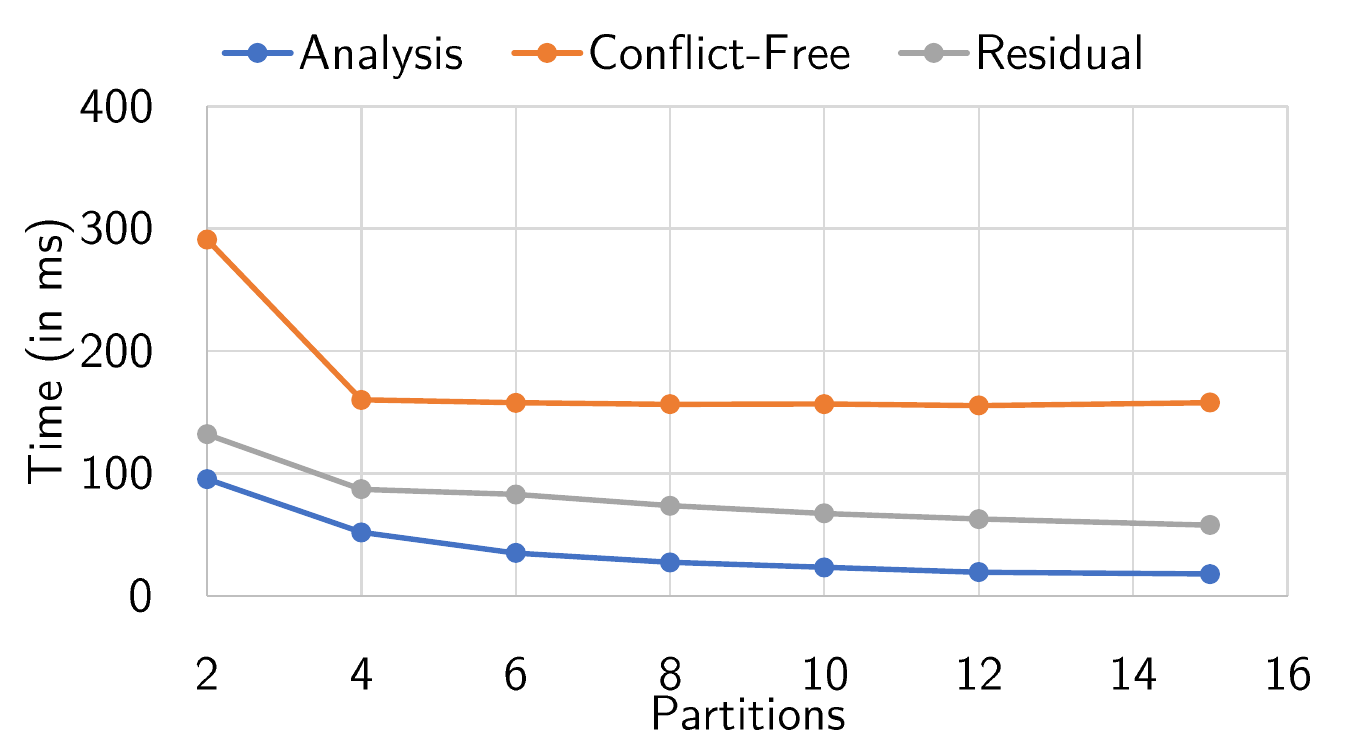}
    \figlabel{tpcc-scalability-pipeline}
    \end{subfigure}
    \caption{Scalability of \tpcc workload with $4$ warehouses: (a) Throughput vs. Cores (b) Runtime Breakdown}
    \figlabel{tpcc-scalability}
\end{figure*}

\begin{figure*}
    \centering
    \begin{subfigure}[t]{0.45\textwidth}
    \includegraphics[width=\linewidth]{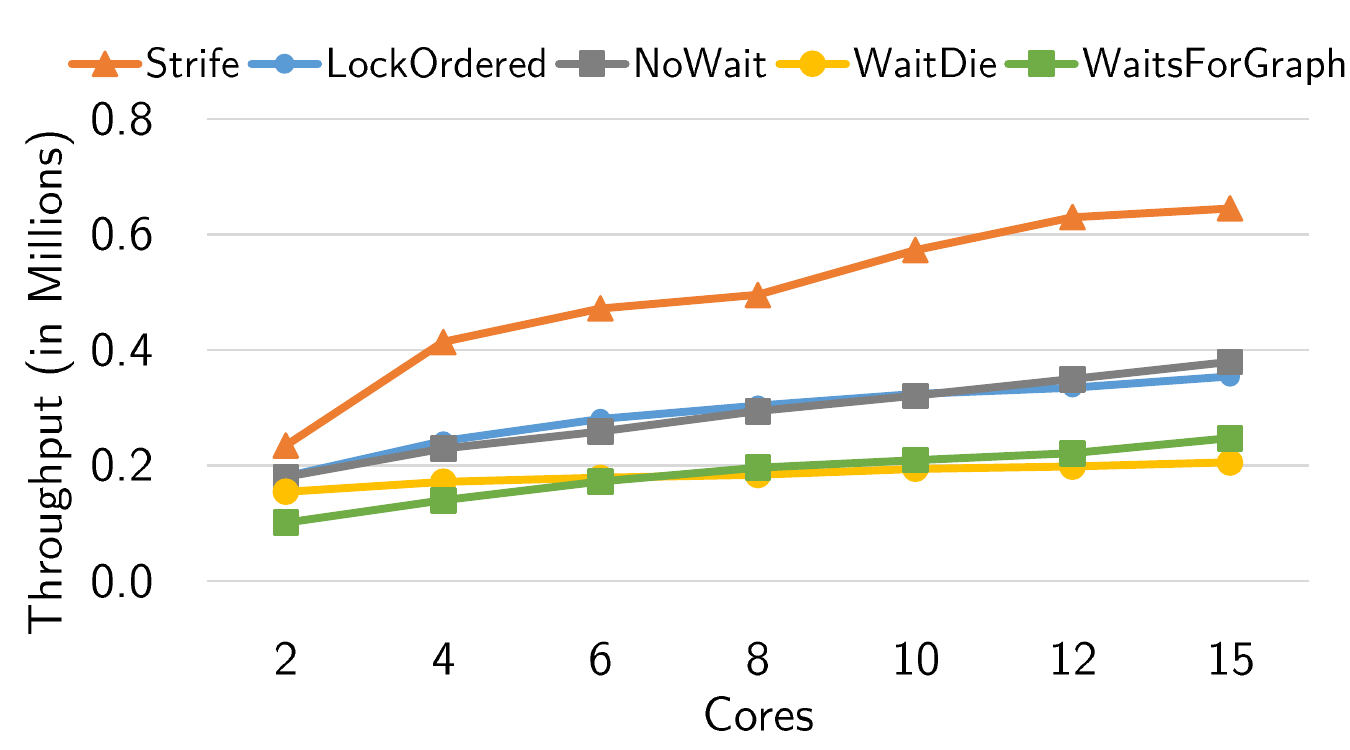}
    \figlabel{ycsb-scalability-main}
    \end{subfigure}
    \hspace{.5cm}
    \begin{subfigure}[t]{0.45\textwidth}
    \includegraphics[width=\linewidth]{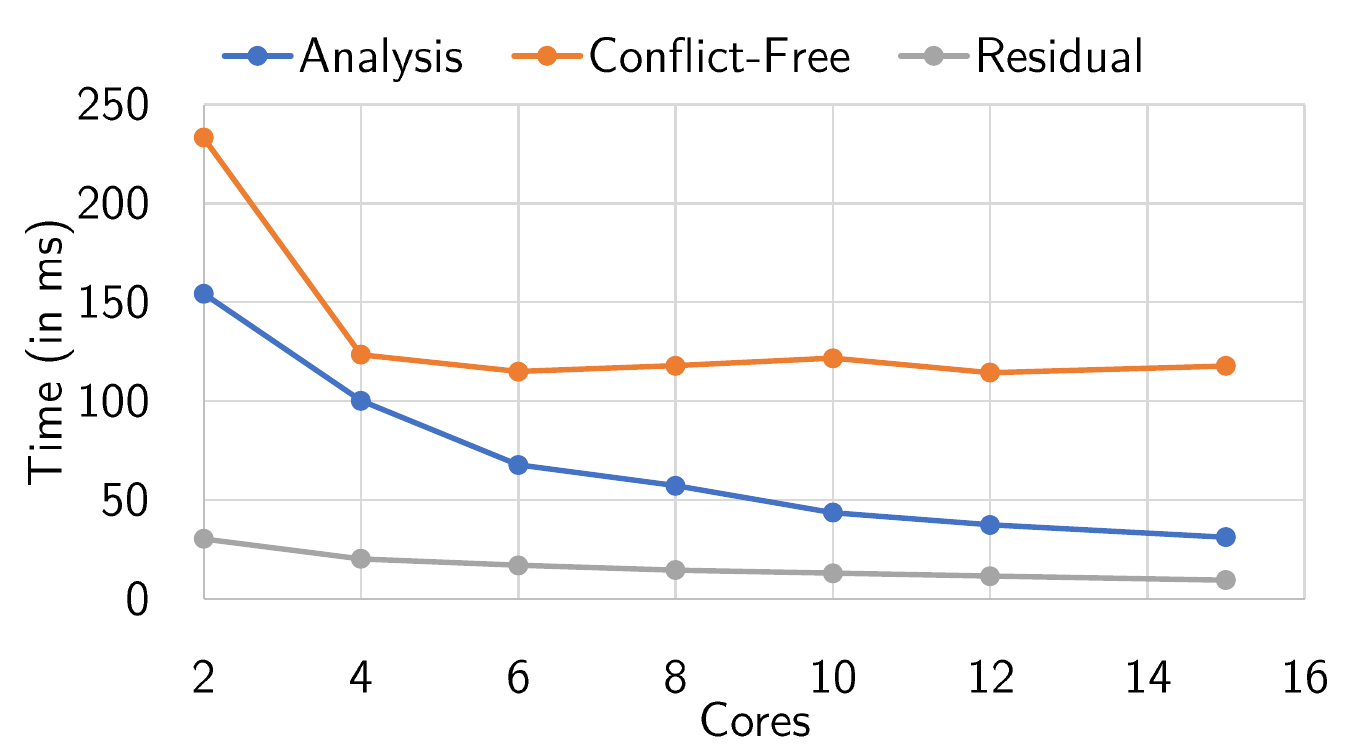}
    \figlabel{ycsb-scalability-pipeline}
    \end{subfigure}
   \caption{Scalability of YCSB workload with $4$ partitions: (a) Throughput vs. Cores (b) Runtime Breakdown}
   \figlabel{ycsb-scalability}
\end{figure*}

The results are shown in \figref{tpcc-scalability}(a). When number of cores is $2$, performance of \system and other protocols are almost similar. \system clusters the batch into $4$ conflict-free clusters and residuals, which are then executed concurrently on $2$ cores. When number of cores is increased to $4$, throughput of \system doubles as all the $4$ clusters can be executed simulatenously. Beyond $4$ cores, the number of conflict-free clusters produced is still $4$ and so there is no significant change in throughput. But since the \analysis phase is executed in parallel, the time spent there decreases, and this improves throughput as we scale up the number of cores.

As number of cores increases, with the same degree of contention (i.e., with $4$ warehouses) other concurrency protocols improve only marginally but are still much poorer than \system. Increasing number of cores in the LockOrdered protocol, for example, results in additional threads unnecessarily spin-waiting on locks. Earlier work has revealed that for very high degrees of parallelism (1024 cores) this can lead to lock thrashing~\cite{dbx1000} and can be more detrimental. 


The WaitsForGraph protocol performs poorly in high contention as the number of transactions added to the waits-for-graph increases as the number of cores increase. Cost of cycle detection increases as it involves acquiring locks on the thread-local state of other threads. In NoWait and WaitDie, on the other hand, more cores result in increased abort rates because the probability that two conflicting transactions access same warehouse concurrently increases. 

For YCSB, we set the number of partitions to be $4$. The keys in the transactions are generated using a zipfian distribution with a $\theta$ value of $0.9$. For $\theta=0.9$, transactions access a set of hot items. \figref{ycsb-scalability}(a) depicts the scalability of our system and the 2PL variants on YCSB workload. Throughput doubles when increasing the cores from $2$ to $4$ due to similar reasons as \tpcc. However, beyond $4$ cores improvement in \system performance is mainly attributed to the parallel execution of \analysis phase.

\figref{tpcc-scalability}(b) and \figref{ycsb-scalability}(b) show the runtime breakdown for \tpcc and YCSB workload respectively. In \tpcc, around $30$\% of time is spent on executing residuals with NoWait concurrency control.  This is due to the remote payments are orders specified in the \tpcc standards specification. Analysis phase in YCSB is more expensive due to the nature of contention. We perform a more detailed analysis in the following section.

\subsection{Factor Analysis}
\sseclabel{factor-analysis}
\begin{figure*}
    \centering
    \begin{subfigure}[t]{0.45\textwidth}
    \includegraphics[width=\linewidth]{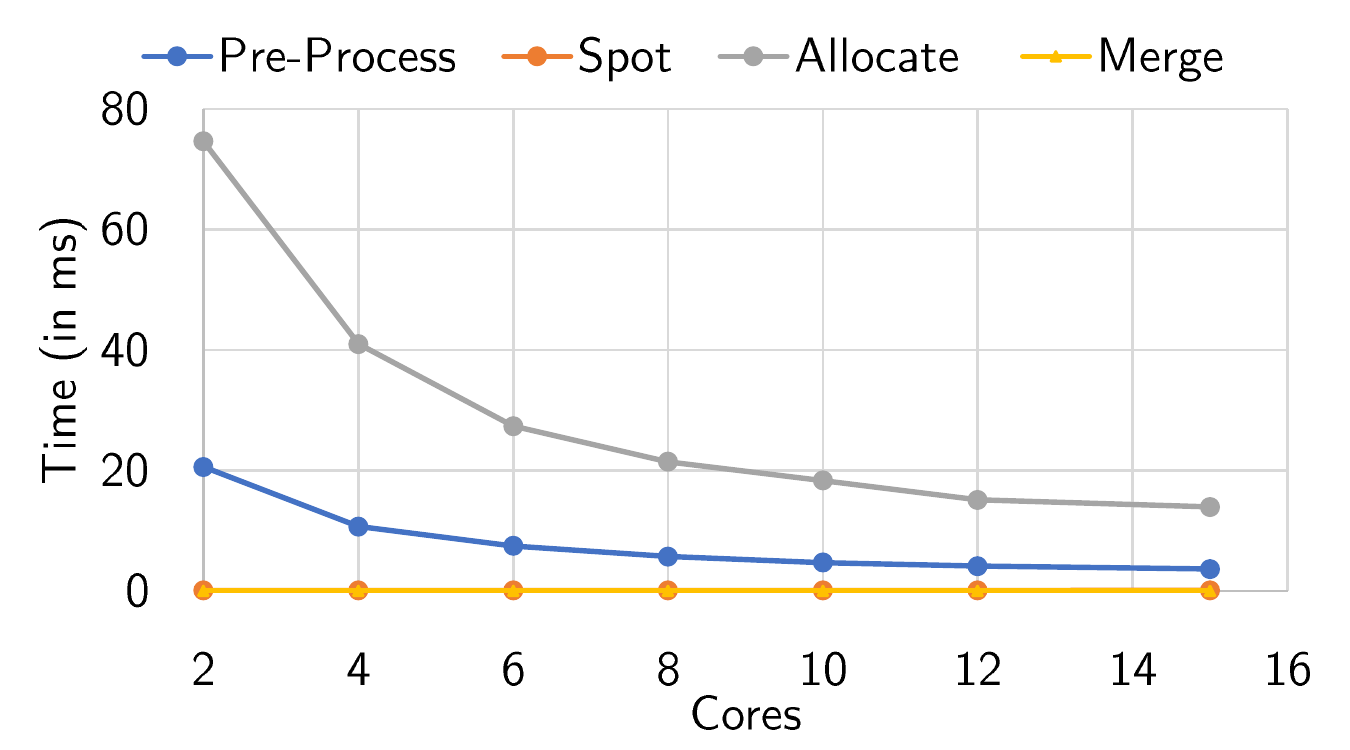}
    \figlabel{tpcc-factor-analysis}
    \end{subfigure}
    \hspace{0.5cm}
    \begin{subfigure}[t]{0.45\textwidth}
    \includegraphics[width=\linewidth]{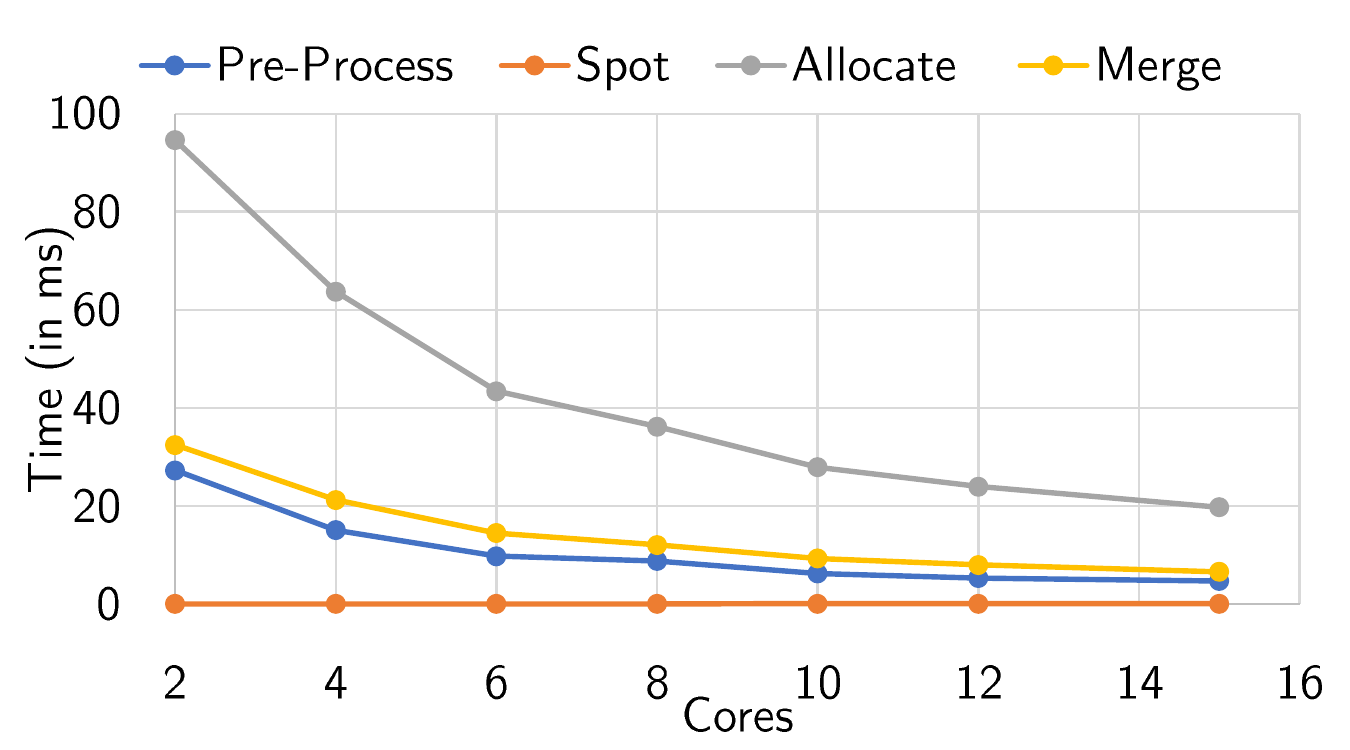}
    \figlabel{ycsb-factor-analysis}
    \end{subfigure}
    \caption{Analysis Phase Breakdown (a) \tpcc with $4$ warehouses (b) YCSB with $4$ partitions}
    \figlabel{factor-analysis}
\end{figure*}

We now analyze the cost of various stages in the \analysis phase. The \analysis phase happens in $4$ stages: \preprocessing, \spot, \allocate and \merge.

\figref{factor-analysis}(a) shows the breakdown of cost of various the stages during analysis of each batch of $100$K transactions from the \tpcc workload with a $50$:$50$ mixture of new-order and payment transactions. With $2$ cores, most of the time (around $80$\%) is spent in allocating transactions to seed clusters and $20$\% on the pre-processing stage. As we increase the degree of parallelism, the time taken by the \allocate and \preprocessing stages drops steadily. Since we sample only a few transactions ($\approx 150$ for our experiments), the cost of \spot stage is almost negligible and hence overlaps the cost of \merge in the figure.

As our \tpcc transactions can be clustered based on the warehouse that they access,
the \spot stage is able to identify seed clusters that correspond to warehouses easily. We set the value of $\alpha$ (that determines ratio of size of residual cluster) to be $0.2$ in all our experiments. Based on the seed clusters, \system is able to allocate all transactions to a cluster corresponding to its warehouse cluster when there are no remote accesses and to the residual cluster when there are remote accesses. The main observation here is that \system does not enter the \merge stage as the number of transactions in \residual cluster is already within the bounds specified by $\alpha$.  Most of the time in \analysis phase is spent in scanning through the read-write sets of transactions to allot them to a cluster in the \allocate stage. The reduction in \allocate and \preprocessing stage cost is reflected marginally in the overall performance, as shown in \figref{tpcc-scalability}(a).

The \analysis phase breakdown for YCSB workload is shown in \figref{factor-analysis}(b). YCSB is different from \tpcc in that it has a set of hot items belonging to each partition. Let $d_1$ and $d_2$ be two such hot items: if $d_1$ and $d_2$ are selected into two different clusters in the \spot stage, then all data items co-accessed with $d_1$ get allocatted separately and those with $d_2$ separately. Essentially it creates a partition within the YCSB partition rendering most transactions as residual in the \allocate stage. Since the residual cluster size is large, we spend more time merging them.

\subsection{Impact of Contention}
\sseclabel{contention}
\begin{figure}
    \centering
    \includegraphics[width=0.9\linewidth]{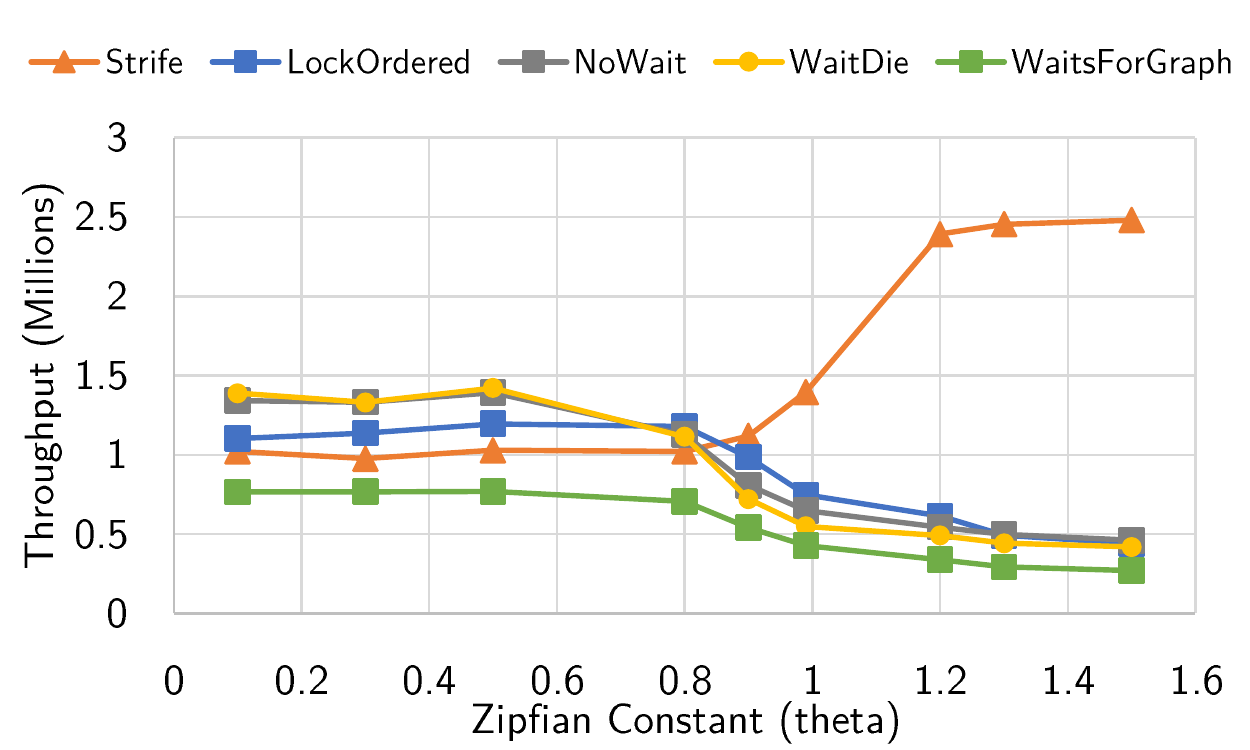}
    \caption{Contention Analysis on YCSB ($15$ partitions)}
    \figlabel{ycsb-contention-theta-main}
\end{figure}

Finally, we compare the performance of \system with other protocols by varying the contention parameter $\theta$ of the YCSB\footnote{A similar experiment is not possible for the \tpcc workload as every new-order and payment transaction accesses the warehouse tuple, making it highly contended within the warehouse.} workload. Even though \system is not designed for low-contention workloads, we present this analysis to empirically understand the behavior of our partitioning algorithm.

The zipfian constant $\theta$ determines the frequency of access to popular items. For a small value of $\theta$ most transactions access different items leading to fewer conflicts in the workload. In the low contention case, most concurrency control protocols perform well. Especially, we see that NoWait and WaitDie protocols perform about $50$\% and LockOrdered about $20$\% better than \system as shown in ~\figref{ycsb-contention-theta-main}. But, when we increase the value of theta, the number of conflicts in the workload increases as many transactions access a few hot items. In this high contention case, \system outperforms other protocols by $5$x. Compared to the overheads of executing such a workload with concurrency control, the \analysis quickly reveals the underlying structure of the batch which is used to schedule conflict-free clusters in parallel without any concurrency control. 

\section{Discussion}
\seclabel{discussion}

In this section we discuss a few tradeoffs in the design of \system. As discussed in~\secref{introduction}, \system is inspired by empirical studies~\cite{oltp-looking-glass, dbx1000} that reported concurrency control to be a substantial overhead for transactional workloads with high contention. It is built on the insight to identify conflict-free clusters in workloads that can be executed in parallel without concurrency control.

The main challenge in realizing this insight is to partition transactions into conflict-free clusters (as discussed in~\secref{architecture}) fast enough that does not outweigh the benefits of concurrency control free parallel execution.
Most traditional approaches to graph partitioning (as discussed in~\secref{related-work}) do not meet this criterion. Hence, we designed a new heuristic that exploits the amount of data contention in each workload. Following are some pros and cons of our design choices:

\begin{myitemize}
\item Randomly sampling transactions (\secref{spot}) allows us to quickly spot contentious items in the workload to form initial seed clusters. Most other techniques require expensive tracking of access patterns for each data item. 
\item Another key observation that is specific to high contention workloads is that diameter of the access graph is often small (close to $1$). We characterize this in~\ssecref{contention}. This allows us to optimize the \allocate phase by only needing to run atmost 2  rounds of allocation. 
\item On the contrary, for low contention workloads, the diameter of the access graph tends to be greater than 1. Hence we either need to run multiple rounds of \allocate, or assign transactions randomly to the initial seed clusters, as detailed in \ssecref{allocate}.
The former increases the amount time spent in \analysis, while the latter results in sub-optimal clusters. We currently choose the latter, although the resulting throughput is still comparable to other locking-based protocols, as shown in~\figref{ycsb-contention-theta-main}.

\item \merge step uses an additional parameter $\alpha$ that determines which clusters to merge. Merging clusters results in two competing effects: (1) it reduces parallelism in the fast \conflictfree phase; but (2) increases the number of transactions that are executed without concurrency control. A high value of $\alpha$ reduces the overall benefit of using \system, while a small value of $\alpha$ can force merging of clusters and reduce parallelism. So, picking the right $\alpha$ is important to achieve good performance.

\item The number of conflict-free clusters produced for a batch is an indicator of reasonable trade-off between performance and core utilization. This quantity can be used to provision the amount of resources for the OLTP component in hybrid transactional/analytical processing (HTAP)~\cite{htap} databases.

\end{myitemize}

\section{Related Work}
\seclabel{related-work}

\paragraph{Graph Partitioning}
The scheduling problem we proposed and provided a heuristic problem can be fundamentally modeled as a graph partitioning problem called the $k$-way min cut optimization. Even though, the partitioning problem is NP-Complete, several algorithms have been developed that produces good partitions: including spectral methods~\cite{spectral-1, spectral-2} and geometric partition methods ~\cite{geometric-partition, geometric-partition-2} among others. Multi-level partitioning algorithms~\cite{multi-level} are known to produce better partitions with a moderate computational complexity -- the basic idea is that a large graph is coarsened down to a few hundred vertices, a bisection of this graph is computed and then projected back to the original graph through multi-step refinement. METIS~\cite{metis} is an open-source graph partitioning library developed based on this scheme. Our preliminary investigation revealed that these techniques are much more expensive and does not match the practical low-latency processing requirements of OLTP systems.

\paragraph{Data Clustering}
Our scheduling problem computes a clustering of data items simultaneously as we cluster transactions to ascertain conflict-freedom among the conflict-free clusters. An alternative approach that we investigated is to first partition the data items based on co-occurrence in the batch followed by clustering of transactions based on this partition. Most of the data clustering algorithms such as $k$-means clustering are iterative. Multiple scans of transactions and its read-write set incurs a significant overhead compared to the actual execution of transactions. However, our randomized solution is inspired by Ailon et. al.~\cite{ailon} solution to the correlation clustering problem where elements are clustered based on a similarity and dissimilarity score.

\paragraph{Lock Contention}
Johnson et. al.~\cite{speculative-lock-inheritance} identify the problem of lock contention on hot items and apply a speculative lock inheritance technique to skip the interaction with a centralized lock table by directly passing over locks from transactions to transactions; a core assumption in this work, which does not apply to the highly contended workloads we deal with, is that transactions mostly acquire shared locks on the hot items. Jung et. al.~\cite{scalable-lock-manager} identify lock manager as a significant bottleneck and propose a new design for lock manager with reduced latching. We have shown that \system outperforms the 2PL protocols even under the optimistic assumption of highly scalable record-level locks. Sadoghi et. al.~\cite{lock-contention-mvcc} identify lock contention as a significant overhead proposes an MVCC based optimization to reduce it.

Orthrus~\cite{high-contention} is a database design proposal for high contention workloads that partition the concurrency control and transaction execution functionalities into different threads. However, unlike our design, Orthrus still uses locks to perform concurrency control. Yu et. al.~\cite{dbx1000} evaluate OLTP performance on 1000 cores and report that locking-based protocols are perform worse on write-intensive workloads due to lock thrashing, while lightweight 2PL protocols such as NoWait and WaitDie result in a very high abort rate.

\paragraph{Modular and Adaptive Concurrency Control}
Callas~\cite{modular-concurrency-control} presents a modular concurrency control to ACID transactions that partitions transactions into groups that when executed independently under different concurrency protocols still ensures serializability. Callas uses the dependency graph of a workload similar to our system to group data items but is different from our approach in that we analyze every batch independently and hence can adapt to changing access patterns quickly.
\texttt{IC3}~\cite{ic3} uses static analysis and dynamic dependency tracking to execute highly contended transactions as pieces on multiple cores in a constrained fashion that ensures serializability. Tang et. al. propose adaptive concurrency control(ACC) that dynamically clusters data items and chooses optimal concurrency control for each cluster using a machine learning model trained offline.

\paragraph{Improvements to Traditional Protocols}
Dashti et. al.~\cite{mvcc-repair} propose a new approach for validating MVCC transactions that uses the dependency graph to avoid unnecessary aborts of transactions. BCC~\cite{bcc} improves traditional OCC by dynamically tracking dependencies that help avoid false aborts during the validation phase.
Yan et. al.~\cite{cong} improve 2PL by statically analyzing stored procedures to find an efficient lock acquisition order based on contention of data items in the workload.

\paragraph{Partitioned and Deterministic Databases}
H-Store~\cite{hstore, hstore-cc} partitions the database such that most transactions access a single partition thereby reducing the overall concurrency control overhead. Partitioned databases requires static determination of partitions and does not adapt to changing access patterns. Moreover, multi-partition transactions are known to cause a significant drop in OLTP performance~\cite{dbx1000} for partitioned databases. Pavlo et. al.~\cite{skew-aware} automatically repartitions a database based on a given workload using local neighborhood search.
Calvin~\cite{calvin} is a distributed database that executes transactions by deterministically ordering them. Our choice of micro-batching transactions to schedule them optimally is inspired Calvin and related literature~\cite{adv-disadv} on  determinstic databases.

Pavlo et. al.~\cite{predictive-modeling} predict and choose the optimizations (such as intelligent scheduling) that a distributed OLTP system can employ during runtime using a combination of offline machine-learning and markov models. This approach, however, is not adaptive to dynamic workloads with changing access patterns.

\section{Conclusion}
\seclabel{conclusion}
We presented \system, a transaction processing system for high-contention workloads. \system is designed based on the insight that portions of a transactional workload can be executed as conflict-free clusters without any concurrency control, even when the workload has high data contention.
We achieved this by developing a low-overhead partitioning algorithm that divides a batch of transacions into a set of conflict-free clusters and residuals.
The clusters are executed on multiple cores without any  concurrency control, followed by the residuals executed with concurrency control. Our experiments have showed that \system can achieve substantial performance improvement, with $2\times$ throughput increase compared to standard locking-based protocols on \tpcc and YCSB workloads.


\bibliographystyle{ACM-Reference-Format}
\bibliography{main}

\end{document}